\documentclass[lettersize,journal]{IEEEtran}
\usepackage{amsmath,amsfonts}
\usepackage{booktabs}
\usepackage{array}
\usepackage{textcomp}
\usepackage{stfloats}
\usepackage{url}
\usepackage{verbatim}
\usepackage{graphicx}
\usepackage{cite}
\usepackage[ruled,linesnumbered]{algorithm2e} 
\usepackage{multirow}
\usepackage{multicol}
\usepackage{graphicx}
\usepackage{comment}
\usepackage{animate}
\usepackage{subcaption}
\usepackage{algpseudocode}
\usepackage{xcolor}
\usepackage{makecell}
\usepackage{cite}
\usepackage[colorlinks=true,linkcolor=blue,citecolor=blue,urlcolor=blue]{hyperref}

\begin{document}

\title{Disentangled 4D Gaussian Splatting: Rendering High-Resolution Dynamic World at 343 FPS}

\author{Hao Feng, Wei Xie, Hao Sun, Zhi Zuo, Zhengzhe Liu
\thanks{This work was supported in part by the National Natural Science Foundation of China under Grant 62377026 and in part by 
the Fundamental Research Funds for the Central Universities under Grant CCNU25JC045.}

\thanks{Hao Feng, Wei Xie, Hao Sun are with Hubei Provincial Key Laboratory of Artificial Intelligence and Smart Learning, Central China Normal University, Wuhan 430079, China, with the School of
Computer Science, Central China Normal University, Wuhan 430079, China,
and also with the National Language Resources Monitoring and Research
Center for Network Media, Central China Normal University, Wuhan 430079,
China (e-mails:xw@mail.ccnu.edu.cn). Zhi Zuo is with College of Artificial Intelligence,
Nanjing University of Aeronautics and Astronautics, Nanjing 211106, China. Zhengzhe Liu  is with School of Data Science,
Lingnan University, Hong Kong SAR, China.}
}

\markboth{Journal of \LaTeX\ Class Files,~Vol.~14, No.~8, August~2021}%
{Shell \MakeLowercase{\textit{et al.}}: A Sample Article Using IEEEtran.cls for IEEE Journals}

\IEEEpubid{0000--0000/00\$00.00~\copyright~2021 IEEE}

\maketitle


\begin{abstract}
While dynamic novel view synthesis from 2D videos has seen progress, achieving efficient reconstruction and rendering of dynamic scenes remains a challenging task. In this paper, we introduce Disentangled 4D Gaussian Splatting (Disentangled4DGS), a novel representation and rendering pipeline that achieves real-time performance without compromising visual fidelity. Disentangled4DGS decouples the temporal and spatial components of 4D Gaussians, avoiding the need for slicing first and four-dimensional matrix calculations in prior methods. By projecting temporal and spatial deformations into dynamic 2D Gaussians and deferring temporal processing, we minimize redundant computations of 4DGS. Our approach also features a gradient-guided flow loss and temporal splitting strategy to reduce artifacts. Experiments demonstrate a significant improvement in rendering speed and quality, achieving 343 FPS when render $1352\times1014$ resolution images on a single RTX 3090 while reducing storage requirements by at least 4.5\%. Our approach sets a new benchmark for dynamic novel view synthesis, outperforming existing methods on both multi-view and monocular dynamic scene datasets.
\end{abstract} 

\begin{figure*}[h!]
    \begin{center}
    \captionsetup{type=figure}
    \begin{subfigure}{0.23\linewidth}
        \includegraphics[width=0.99\linewidth,height=0.7\linewidth]{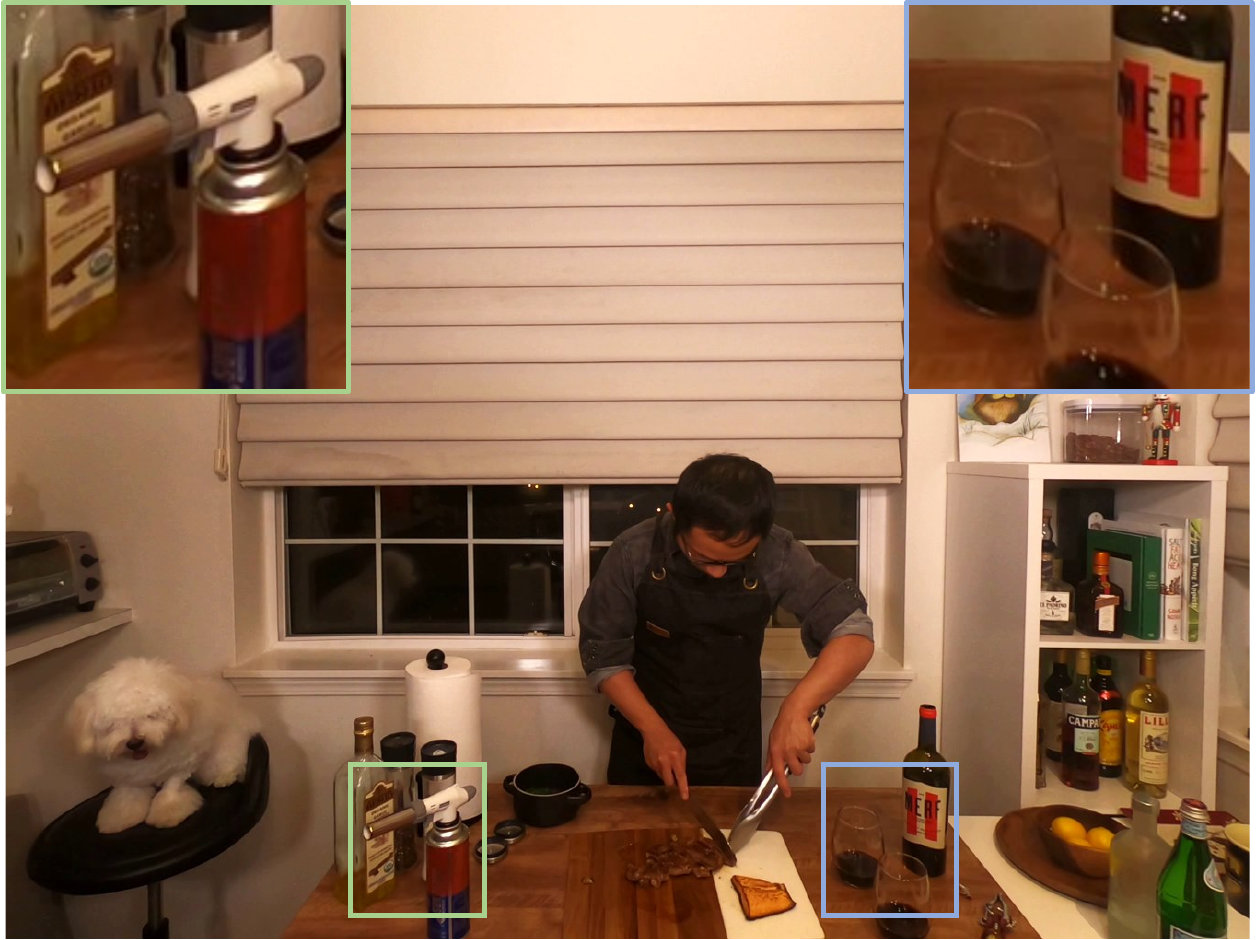}
    \caption{Ground Truth}
    \label{}
    \end{subfigure}
    \centering
    \begin{subfigure}{0.23\linewidth}
        \includegraphics[width=0.99\linewidth,height=0.7\linewidth]{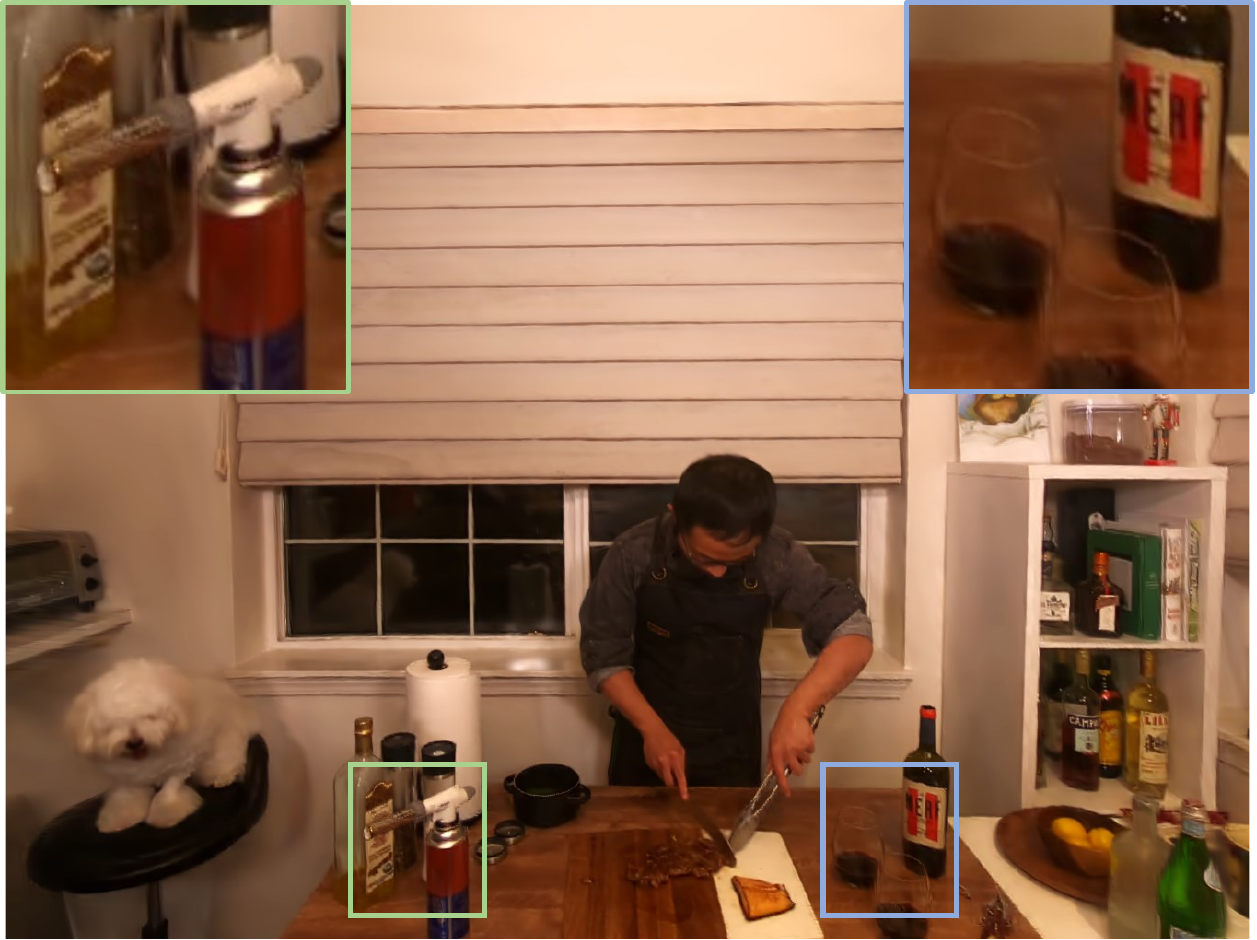}
    \caption{RealTime4DGS\cite{Yang2023RealtimePD}}
    \label{}
    \end{subfigure}
    \centering
    \begin{subfigure}{0.23\linewidth}
        \includegraphics[width=0.99\linewidth,height=0.7\linewidth]{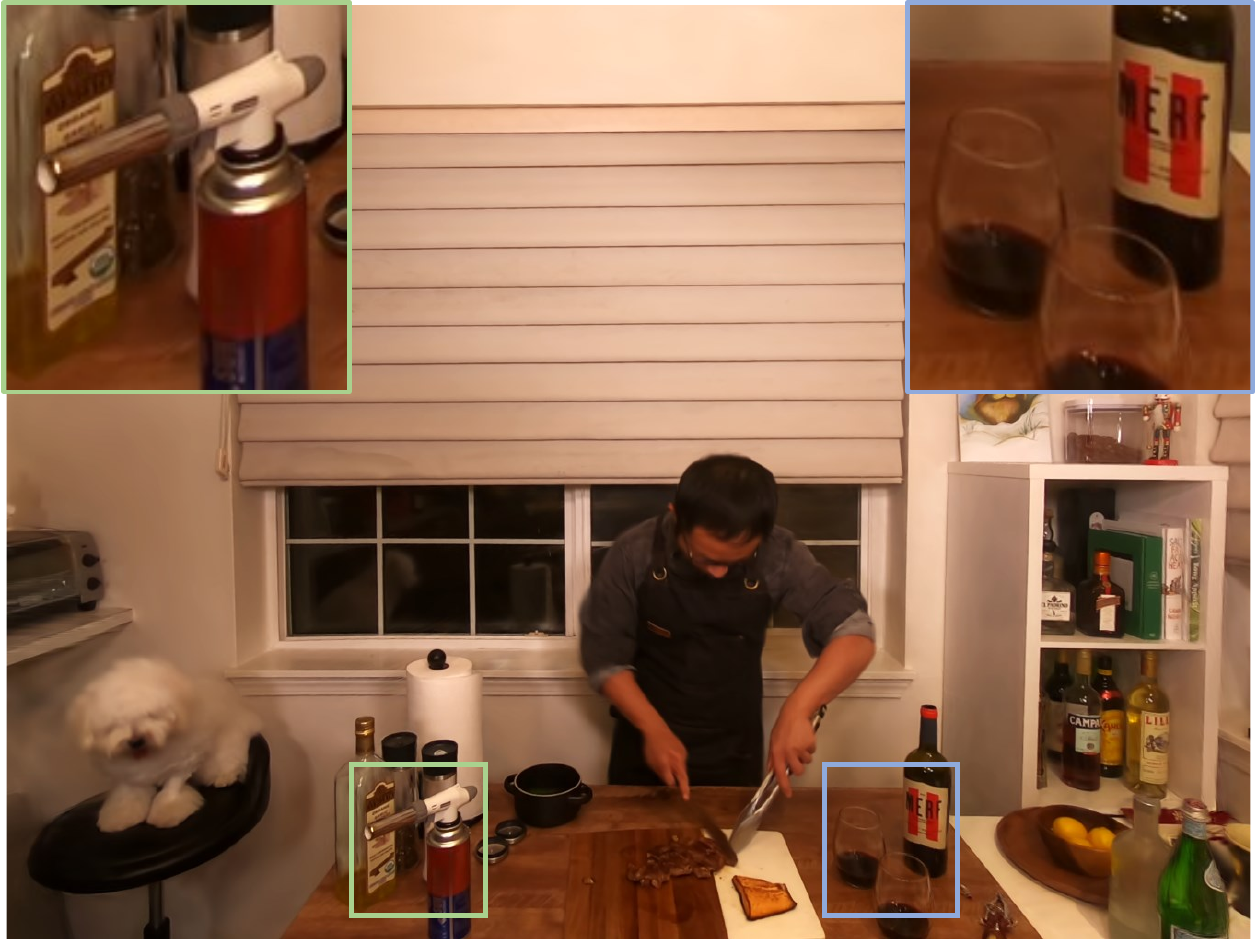}
    \caption{The Proposed Method}
    \label{}
    \end{subfigure}
    \centering
    \begin{subfigure}{0.23\linewidth}
        \includegraphics[width=0.99\linewidth,height=0.7\linewidth]{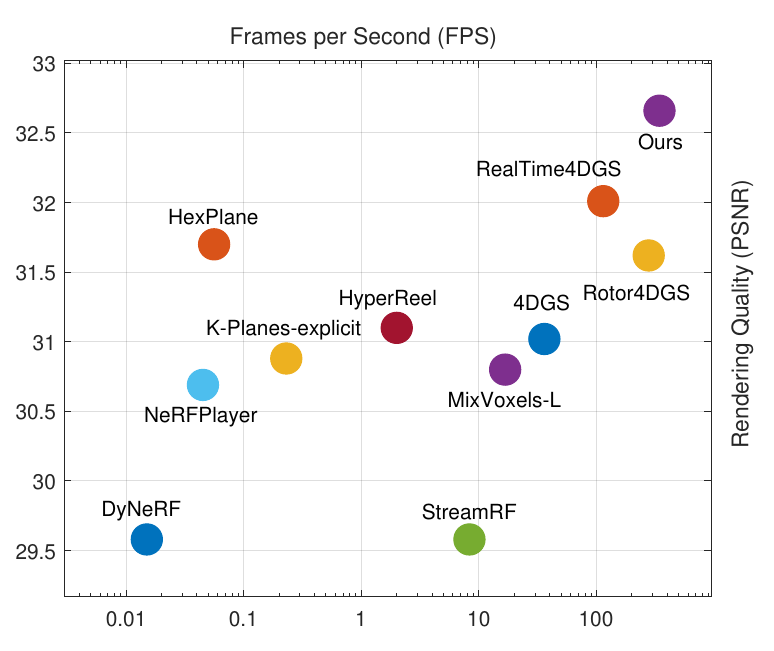}
    \caption{Evaluation on PSNR \textit{vs.} FPS}
    \label{}
    \end{subfigure}
    \setcounter{figure}{0}
    \captionof{figure}{We present Disentangled 4D Gaussian Splatting, a highly-efficient approach that renders 1352 × 1014 resolution images at 343 FPS on an RTX 3090 in the Plenoptic Dataset\cite{9878989}, surpassing previous approaches in both rendering quality and speed. Note that the x-axis is logarithmic scale.} 
    \label{Fig 1.}
\end{center}
\end{figure*}
\section{Introduction}

Reconstructing dynamic scenes from 2D images and synthesizing photo-realistic novel views in real-time, has been a long-standing goal in computer vision and graphics.  This task has attracted increasing attention from both academia and industry because of its potential value in various applications, including film, gaming, and VR/AR\cite{10168294}. While promising progress has been made, achieving both high-quality and highly efficient rendering for dynamic scenes remains a substantial challenge. Building upon the success of 3DGS in static settings, researchers\cite{Yang2023RealtimePD},\cite{Wu20234DGS},\cite{Duan20244DRotorGS} have begun to explore its generalization to the spatio-temporal (4D) domain, aiming to achieve real-time rendering for dynamic  scenes. 

\begin{figure*}
    \centering
            \centering
            \begin{subfigure}{0.9\linewidth}
                \includegraphics[width=0.99\linewidth]{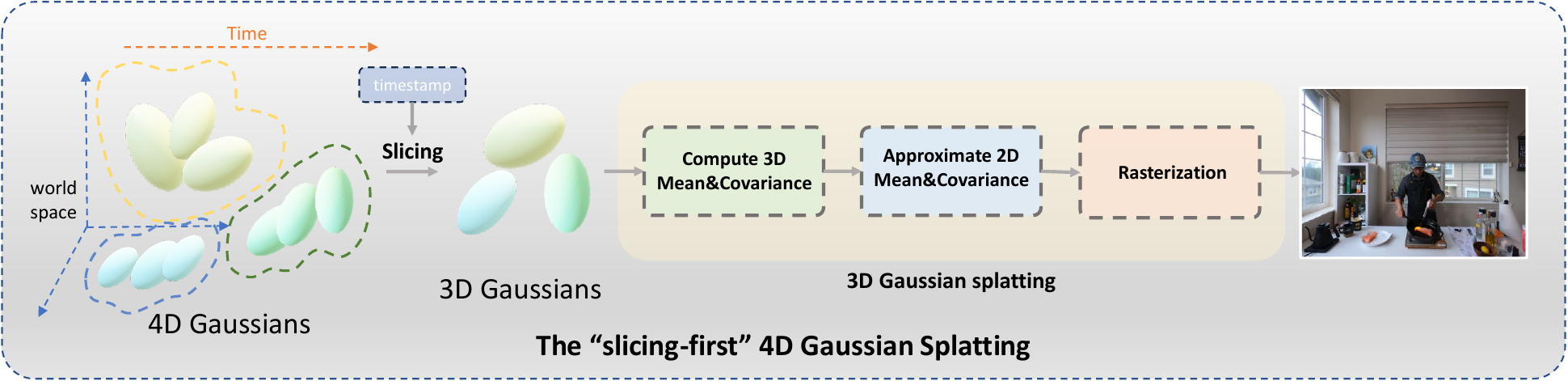}
            \caption{The "slicing-first" 4D Gaussian Splatting}
            \label{Fig 2.(a)}
    \end{subfigure}
            \centering
            \begin{subfigure}{0.9\linewidth}
                \includegraphics[width=0.99\linewidth]{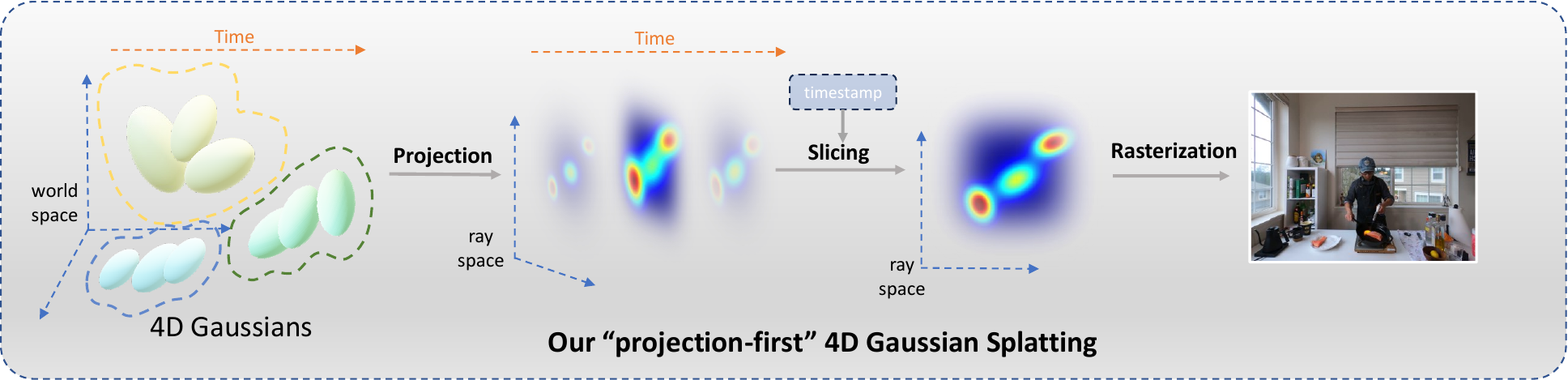}
            \caption{Our disentangled 4D Gaussian Splatting}
            \label{Fig 2.(b)}
    \end{subfigure}
    \caption{Comparison between "slicing-first" 4D Gaussian Splatting and our Disentangled 4D Gaussian Splatting. The upper one is the slicing first 4D Gaussian Splatting method, which need to slice the 4D Gaussian into 3D Gaussian. This approach requires computing high-dimensional covariance matrices and performing repeated slicing and projection operations, leading to inefficiency and temporal discontinuity. In contrast, our "projection-first" disentangled formulation preserves temporal information throughout the rendering pipeline, enabling efficient rasterization and continuous, temporally coherent image synthesis.}
    \label{fig:enter-label}
\end{figure*}

However, existing 4DGS methods face significant efficiency limitations. Existing works on 4DGS can be categorized into two branches. The first branch \cite{Yang2023RealtimePD},\cite{Wu20234DGS} employs deep learning to implicitly model temporal changes in 3D Gaussian ellipsoids. These methods predict variations to the base 3D Gaussian ellipsoids and synthesize novel views by projecting them into 2D Gaussian ellipses in ray space\cite{1021576}. However, per-Gaussian network queries can be computationally expensive and become a bottleneck for real-time performance in large scenes.
The second branch \cite{Duan20244DRotorGS},\cite{Yang2023RealtimePD} extends 3D Gaussian ellipsoids into 4D Gaussian hyperspheres, modeling dynamic scenes through four-dimensional transformations of rotation, scaling, and mean. As shown in Fig.\ref{Fig 2.(a)}, to generate novel views at specific timestamps, these methods first temporally slice the 4D Gaussian hypersphere into a 3D Gaussian ellipsoid corresponding to the given timestamp before proceeding with the 3DGS rendering pipeline. 
Despite of their differences in design, a key limitation shared by both approaches is that they must process (or generate) a static 3D scene representation for each specific timestamp before the main rendering steps. 
Specifically, in the first branch, the implicit deep learning model does not support direct dynamic projection. In the second branch, temporal transformations embedded within the 4D covariance matrix make direct projection computationally expensive and complex\cite{1021576}.  Consequently, these methods must repeat the entire rendering process whenever the timestamp changes, leading to significant redundant computations.

To address the inefficiency caused by redundant computations in existing 4DGS methods\cite{Duan20244DRotorGS},\cite{Yang2023RealtimePD}, we propose Disentangled 4D Gaussian Splatting (Disentangled4DGS). Our core innovation lies in a novel 4D Gaussian representation that disentangles time-dependent properties from other parameters, such as 3D Gaussian parameters, time scaling, and the velocity of the mean. This disentangled representation unlocks a more efficient rendering pipeline. Unlike existing "slicing-first" methods (Fig. \ref{Fig 2.(a)}) that rely on slicing the 4D Gaussian at each timestamp, bring redundant matrix operations with expensive computation, our method (Fig. \ref{Fig 2.(b)}) allows for timestamp updates to directly affect the projected 2D Gaussians, eliminating the need to repeatedly execute the entire 4D-to-2D pipeline repeatedly. Benefiting from our uniquely designed Disentangled 4DGS, our method is memory-efficient and achieves substantial computational speedups by deferring temporal processing and avoiding complex 4D matrix operations.
\IEEEpubidadjcol

Another challenge of 4DGS lies in accurately modeling object motions in complex dynamic scenes. In practice, object trajectories may become distorted, especially near motion boundaries, due to the lack of explicit geometric constraints on dynamic regions. To address this issue, we introduce a flow-gradient guided consistency loss that leverages image gradients as structural cues to regulate motion boundaries. By aligning the gradients of motion fields with image edges, this loss encourages that motion discontinuities occur only at true object boundaries while remaining smooth within homogeneous regions. Our experiments demonstrate that this constraint leads to more coherent object trajectories and improves rendering quality in dynamic scenes.

Furthermore, existing 4DGS approaches\cite{Li2023SpacetimeGF,Duan20244DRotorGS,Yang2023RealtimePD} often approximate object motions with first-order or second-order models. While such simplifications suffice for relatively smooth trajectories, they struggle with complex and non-linear dynamics such as sudden appearance, disappearance, or occlusion changes. To overcome this limitation, we introduce temporal splitting strategies that explicitly decouple complex motions into finer temporal segments, enabling a more accurate representation of non-linear object trajectories. Experimental results confirm that combining flow-gradient guided consistency with temporal splitting significantly enhances motion fidelity and overall rendering quality in dynamic environments.

In summary, the key contributions of Disentangled4DGS are:
\begin{itemize}
    \item We propose a disentangled representation for 4D Gaussians that separates time-dependent components from other parameters, largely improving the efficiency of 4DGS. 
    \item We introduce a novel rendering pipeline for 4D Gaussians that postpones the slicing process, effectively minimizing redundant computations in dynamic scene rendering. 
    \item We propose a flow-gradient guided consistency loss and a temporal splitting strategy to better constrain object motions and improve rendering quality in dynamic scenes. 
    \item Disentangled4DGS achieves unprecedented rendering performance, generating \begin{math}
    1353\times1014
\end{math} resolution videos at an average of \textbf{343 FPS} on an RTX 3090 in Plenoptic Video Dataset \cite{9878989}, as shown in Fig.\ref{Fig 1.}. Both quantitative and qualitative evaluations demonstrate its superiority over previous methods, setting new benchmarks in rendering quality and efficiency. 
\end{itemize}  
\label{sec:intro}

\section{Related work}
\subsection{Novel view synthesis for static scenes}
In recent years, novel view synthesis has received widespread attention. Prior work formalized concepts like lumigraph \cite{Buehler2001UnstructuredLR} or light-field \cite{Levoy2023LightFR} and generating novel-view images through interpolating the existing views. Although traditional methods are efficient, they require densely captured images in complex scenes. Ben et al. \cite{Mildenhall2020NeRF} initiated a significant trend by utilizing a Multi-Layer Perceptron (MLP) to learn the radiance field and applying volumetric rendering for photo-realistic image synthesis from any viewpoint. Subsequent efforts have aimed to accelerate training and rendering \cite{DBLP:journals/corr/abs-2111-12077},\cite{Chen2022TensoRFTR},\cite{Yu2021PlenoxelsRF},\cite{Mller2022InstantNG},\cite{Wan2023LearningND},\cite{10204053} and enhance rendering quality by addressing existing issues in the NeRF \cite{Mildenhall2020NeRF}, such as aliasing and reflection. However, NeRF \cite{Mildenhall2020NeRF} based methods involve querying the MLP for hundreds of points per ray, significantly consuming time. In contrast, Kerbl et al.'s 3D Gaussian Splatting (3DGS) \cite{kerbl3Dgaussians} offers a novel framework enabling real-time, high-fidelity novel view synthesis for complex scenes. 
Recent advancements  \cite{Yu2023MipSplattingA3},\cite{11189234},\cite{liu20243dgs},\cite{10655416},\cite{Zhang2024CoRGSS3},\cite{Wan2025S2GaussianSS},\cite{11095131} have focused on enhancing both rendering speed, sparse inputs and the quality of synthesized images. 
\subsection{Novel view synthesis for dynamic scenes}
Unlike static scenes, novel view synthesis for dynamic scenes necessitates accurate modeling of geometric shapes and colors, as well as their temporal changes\cite{10.1145/2766945},\cite{7780873},\cite{580394},\cite{10.1145/1015706.1015766}. Inspired by NeRF's success with static scenes, some works \cite{Mller2022InstantNG},\cite{Pumarola2020DNeRFNR} have attempted to extend NeRF \cite{Mildenhall2020NeRF} for dynamic environments, focusing on improving volume and spatiotemporal encoding for efficient dynamic scene modeling. The real-time rendering capabilities of 3DGS for photo-realistic images have drawn widespread attention. Some research \cite{Wu20234DGS},\cite{Kratimenos2023DynMFNM},\cite{10.5555/3692070.3694113},\cite{Chen2025DASH4H},\cite{10.5555/3737916.3741850},\cite{Gao20257DGSUS},\cite{10932755} focuses on modeling temporal changes in 3DGS \cite{kerbl3Dgaussians} using deep learning, yielding high-quality novel view synthesis. However, since each Gaussian sphere necessitates querying the model to capture temporal changes, the rendering speed is slower than the original 3DGS. 
\subsection{Dynamic 3D Gaussians}

There has been significant progress in extending 3DGS to dynamic scenes \cite{Duan20244DRotorGS},\cite{Yang2023RealtimePD}. Yang et al. \cite{Yang2023RealtimePD} characterize 4D Gaussian hyperspheres by expanding the quaternions and 3D scaling vectors of 3D Gaussian spheres into dual quaternions and 4D scaling vectors, and they innovatively use four-dimensional spherical harmonics to capture temporal and spatial color variations.  Duan et al. \cite{Duan20244DRotorGS} employ a spatiotemporal decoupled 16-dimensional vector to represent rotation in four-dimensional space, reducing degrees of freedom and optimization complexity. These methods render images faster than deep learning-based approaches.  But all of these methods need to get a static scene for each timestamp before following the 3DGS rendering pipeline. Thus these methods have low efficiency when timestamp changes frequently, which often occurs in dynamic scenarios. In contrast, by projecting  the Gaussian function in ray space firstly, we can delay the time dimension in rendering, thereby reducing duplicate calculations when time changes.
\section{Method}
\begin{figure*}
    \centering
    \includegraphics[width=1\linewidth]{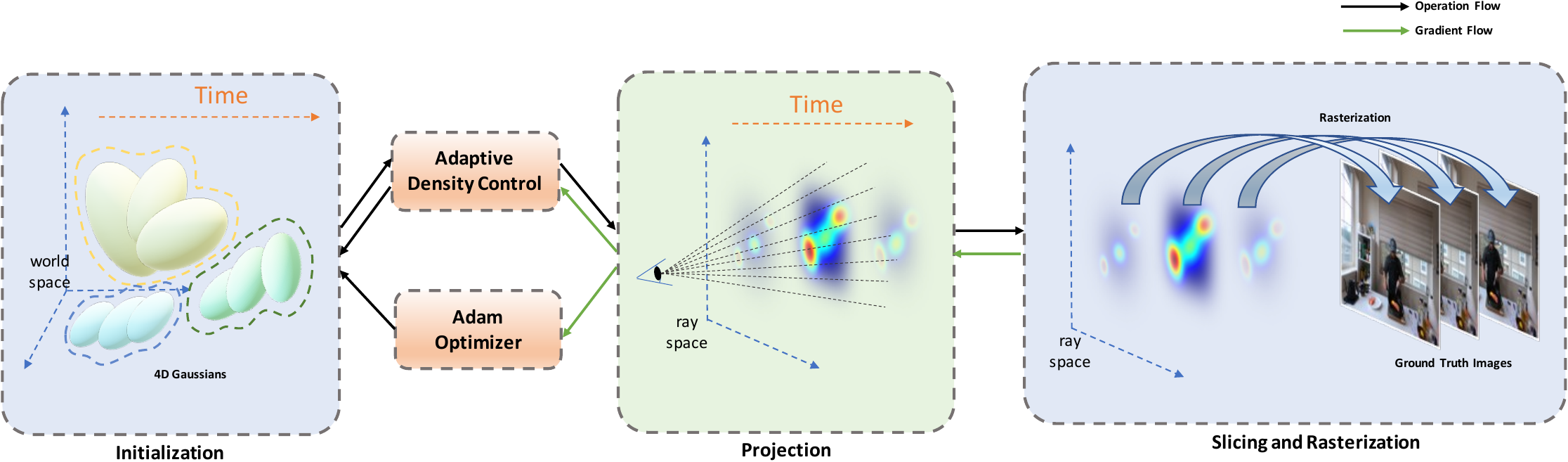}
    \caption{Rendering pipeline of our Disentangled 4DGS. After initialization, we first project the 3D Gaussians and the velocity of mean orthogonally to the timeline, obtaining a 2D Gaussian sphere with velocity in ray space. Then the projected 2D Gaussians with velocity are sliced to obtain the static 2D Gaussian in ray space and utilize rasterization to produce the image. The gradients from loss are back-propagated to optimize the 4D Gaussians and guide the adaptive density control.}
    \label{Fig 3.}
\end{figure*}

In this section, we first review the 3D Gaussian Splatting (3DGS) method \cite{kerbl3Dgaussians}, which inspired our method, in Section 3.1. In Section 3.2, we detail how our method renders the novel view of dynamic scene represented by 4D Gaussian. In Section  3.3, we introduce our spatial edge loss and temporal split strategy. An overview of our framework is shown in Fig.\ref{Fig 3.}.
\subsection{Preliminary of 3D Gaussian Splatting}

3D Gaussian Splatting(3DGS) \cite{kerbl3Dgaussians} has demonstrated real-time photo-realistic rendering capabilities in static scenes. It employs a set of anisotropic Gaussian ellipsoids to represent the scene. Each Gaussian ellipsoid can be characterize by a three-dimensional covariance matrix \begin{math}\Sigma\end{math} and a three-dimensional mean vector \begin{math}\mu\end{math}, as described below:
\begin{equation}
G(x)=e^{-\frac{1}{2}(x-\mu)^{T}\Sigma^{-1}(x-\mu)}
\end{equation}
The initial process of 3DGS rendering is computing the covariance matrix \begin{math}
    \Sigma
\end{math} through scale $s_{3D}$ and quaternion $q$. A view matrix then is used to transform the Gaussian ellipsoids from world space to camera space. Because this transformation is an affine transformation, the  ellipsoids are still Gaussian ellipsoids. To get the 2D reconstruction kernel in ray space \cite{1021576}, we need to perform a mapping \begin{math}
    x=\phi(\mathbf{P_{3D}})
\end{math} on Gaussian ellipsoids, which can be formulated as:\begin{equation}
    \begin{pmatrix}
    x_{0} \\y_{0} \\z_{0}
\end{pmatrix} = \phi(\mathbf{P}_{3D}) =\begin{pmatrix}P_{0}/P_{2} \\P_{1}/P_{2} \\ \left \| (P_{0},P_{1},P_{2})^{T} \right \| 
\end{pmatrix}
\end{equation}
where \begin{math}
    \mathbf{P_{3D}}
\end{math} is the coordinate in camera space and \begin{math}
    x
\end{math} is the coordinate in ray space. To maintain affine transformation properties, ensuring that Gaussian properties are unaltered \cite{1021576}, 3DGS applies the the first two terms of the Taylor expansion of \begin{math}
    \phi
\end{math} to approximation, which can be defined as: \begin{equation}
    \phi_{k}(\mathbf{t})=\mathbf{x}_{k}+\mathbf{J}_{k}(\mathbf{P}-\mathbf{P}_{k})
\end{equation}
where \begin{math}
    \mathbf{J}_{k}
\end{math} is the Jacobian matrix of \begin{math}
    \phi
\end{math} at the point \begin{math}
    \mathbf{t}_{k}
 \end{math}. 
In the end, the color of each pixel in the image is calculated by blending Gaussians sorted by their depths:\begin{equation}
    C=\sum_{i=1}^{N}c_{i}\alpha_{i} \prod_{j=i}^{i-1}(1-\alpha_{j}) 
\end{equation}
where \begin{math}
    N
\end{math} is the number of Gaussians which influence the pixel, \begin{math}
    c_{i}
\end{math} is the color of i-th Gaussian, \begin{math}
    \alpha_{i}=o_{i}G^{'}_{i}
\end{math}. For more details, please refer to \cite{kerbl3Dgaussians,1021576}.
\subsection{Disentangled 4D Gaussian Splatting}
We now introduce our \textbf{Disentangled 4D Gaussian Splatting (Disentangled4DGS)} algorithm, as illustrated in Fig. \ref{Fig 3.}. Specifically, we model the 4D Gaussian using 3D Gaussian (Sec. 3.2.1), time scaling, and velocity of mean. The presentation is then projected into ray space \cite{1021576} (Sec. 3.2.2). We then slice the Gaussian in ray space and apply the fast rasterization technique to render images.
\subsubsection{Representation of 4D Gaussian}
Previous works represent the 4D Gaussian hypersphere using a covariance  matrix \begin{math}
    \Sigma_{4D}
\end{math} and a mean vector  \begin{math}
    \mathbf{\mu}_{4D}=(\mu_{x},\mu_{y},\mu_{z},\mu_{t})
\end{math} to character the shape and position. To ensure that $\Sigma_{4D}$ remains a positive definite symmetric matrix,  they incorporate 4D rotation \begin{math}
    \mathbf{R}_{4D}
\end{math} and 4D scaling \begin{math}
    \mathbf{S}_{4D}
\end{math} into \begin{math}
    \Sigma_{4D}
\end{math} as follows: \begin{equation}
    \Sigma_{4D}=\mathbf{R}_{4D}\mathbf{S}_{4D}\mathbf{S}^{T}_{4D}\mathbf{R}^{T}_{4D}=\begin{pmatrix}
  \mathbf{U}&\mathbf{V} \\
  \mathbf{V^{T}}&\mathbf{W}
\end{pmatrix},
\end{equation}
where $\mathbf{U}$ is the covariance matrix of base 3D Gaussian, $\mathbf{V}$ is the Spatiotemporal covariance, $\mathbf{W}$ is the temporal variance. Extending the 3D scaling matrix to 4D is straightforward; however, extending quaternion-based 3D rotation to 4D is more complex due to the intrinsic coupling between spatial and temporal rotations in the 4D domain. Yang et al. \cite{Yang2023RealtimePD} employs the dual quaternion and Duan et al. \cite{Duan20244DRotorGS} adopt the 4D rotor to characterize the 4D rotations. Both of their slicing methods are based on conditional probability derivations, formulated as: 
\begin{equation}
    G_{3D}(\mathbf{x},t)=e^{-\frac{1}{2}\lambda(x-\mu_{t})^{2}}e^{-\frac{1}{2}[\mathbf{t}-\mathbf{\mu}(t)]^{T}\Sigma^{-1}_{3D}[x-\mathbf{\mu}(t)]},
\end{equation}
where
\begin{equation}
    \begin{aligned}
\label{equation.9}
      \lambda&=\mathbf{W}^{-1},\\
    \Sigma_{3D}&=\mathbf{A}^{-1}=\mathbf{U}-\frac{\mathbf{V}\mathbf{V}^{T}}{\mathbf{W}},\\
    \mu_{3D}&=(\mu_{x},\mu_{y},\mu_{z})^{T}+(t-\mu_{t})\frac{\mathbf{V}}{\mathbf{W}}.
\end{aligned}
\end{equation}
As shown in Eq.~\ref{equation.9},
the resulting 3D mean $\mu_{3D}$ is dependent on the components of the 4D covariance matrix $\Sigma_{4D}$ along the time dimension, even though the 3D mean itself is defined only in spatial coordinates. This inherent spatial-temporal coupling made the direct projection of the combined scale $\mathbf{s}_{4D}$ and rotation $\mathbf{R}_{4D}$ both challenging and computationally expensive. Besides, the projection of 4D covariance matrix $\Sigma_{4D}$ is not an affine transformation, thus a complex Jacobian determinant needs to be calculated to approximate the projection transformation\cite{1021576}.  To address this difficulty, we propose a decomposition of the terms in Eq.~\ref{equation.9}, which yields more physically interpretable components:
\begin{itemize}
    \item \textbf{velocity of mean} $\mathcal{V}_{3D}=\frac{\mathbf{V}}{\mathbf{W}}$,  representing the rate of change of the 3D mean over time.
    \item \textbf{3D Gaussian representation}, where $\mathbf{U}-\frac{\mathbf{V}\mathbf{V}^{T}}{\mathbf{W}}$ serves as covariance matrix $\Sigma_{3D}$, and $(\mu_{x},\mu_{y},\mu_{z})^{T}$ serves as the mean vector $\mu_{3d}$
    \item \textbf{Temporal scaling} $s_{t}=\sqrt{\mathbf{W}}$ is derived from the temporal variance term $\mathbf{W}$ in the 4D covariance matrix $\Sigma_{4D}$, which controls the variance along the time dimension and ensures it remains positive.
\end{itemize}
A detailed proof demonstrating the equivalence of these two representations is provided in the \textbf{supplementary material}. A key advantage of our representation is that the time-dependent variables (velocity of mean $\mathcal{V}_{3D}$, temporal scaling $s_{t}$) are represented as vectors or scalars, which can be quickly and easily projected. 

Therefore, our Disentangled4DGS can be formulated as:
\begin{equation}
\begin{aligned}
    \mathbf{s}_{4D}=&\{\mathbf{s}_{x},\mathbf{s}_{y},\mathbf{s}_{z},\mathbf{s}_{t}\},\\
    \mu_{4D}=&\{\mu_{x},\mu_{y},\mu_{z},\mu_{t}\},\\
    q=&\{q_{1},q_{2},q_{3},q_{4}\},\\
    \mathcal{V}_{3D}=&\{v_{x},v_{y},v_{z}\},\\
\end{aligned}
\end{equation}
where $\mathbf{s}_{4D}$ is the scaling vector, $q$ represents the quaternion, and $\mathcal{V}_{3D}$ denotes the velocity of mean. Our representation effectively disentangles temporal deformations, enabling for efficient and fast projection into the ray space \cite{1021576}. Compared to traditional methods requiring 16 floating-point numbers, our representation only requires 15, thereby improving storage requirements by at least 4.5\%.
\subsubsection{The Differentiable "Projection-First" Splatting for 4D Gaussian}
To enable differentiable rendering of dynamic Gaussians, we project each 4D Gaussian from world space to camera ray space.
Specifically, we compute the mean, velocity, Jacobian, and covariance in ray space. First,the mean of a 4D Gaussian, denoted as $\mu_{4d}=(\mu_{3d},t)$, is tranformed from world space to camera space before projection, which can be expressed as:
\begin{equation}
    P_{3D}=\varphi(\mu_{3D}),P_{t}=\mu_t,
    \end{equation}
    where
    \begin{equation}  
        \varphi(\mu)=\mathbf{W}_\text{view}\mu+d_\text{view}.
\end{equation}
Second, the velocity of mean in camera space $\mathcal{V}_{view}$ can be obtained as:
\begin{equation}
\begin{aligned}
    \mathcal{V}_\text{view }=\varphi^{'}(\mathcal{V}_{3D}),\\
    \varphi^{'}(\mathcal{V})=\mathbf{W}_\text{view}\mathcal{V}
    \end{aligned}
\end{equation}
The projective transformation of the mean vector can be formulated as:
\begin{equation}
   \begin{pmatrix}
 x_{0}\\
 y_{0}\\
 z_{0}
\end{pmatrix}=\phi_{3D}(\mathbf{P}_{3D})= \begin{pmatrix}
 P_{0}/P_{2}\\
 P_{1}/P_{2}\\
 \left\|(P_{0},P_{1},P_{2})^{T}\right\|
\end{pmatrix}, t_{0}= P_{t}
\end{equation}
where $(x_{0},y_{0},z_{0},t_{0})^{T}$ represents the mean in ray space \cite{1021576}. And the velocity of mean in ray space can be easily obtained as:
\begin{equation}
   \begin{pmatrix}
 v_{0}\\
 v_{1}\\
 v_{2}
\end{pmatrix}=\phi^{'}_{3D}(\mathcal{V}_\text{view})= \begin{pmatrix}
 \mathcal{V}_{\text{view}_{x}}/P_{2}\\
\mathcal{V}_{\text{view}_{y}}/P_{2}\\
 \left\|( \mathcal{V}_{\text{view}_{x}},\mathcal{V}_{\text{view}_{y}},\mathcal{V}_{\text{view}_{z}})^{T}\right\|
\end{pmatrix}.
\end{equation}
At last, to derive the covariance matrix $\Sigma^{'}$ in ray space, we compute the Jacobian matrix $\mathbf{J}_{k}$ of $\phi_{3D}$. Following 3DGS, Jk can be formulated as:
\begin{equation}\begin{aligned}
    \mathbf{J}_{k}=\frac{\partial\phi}{\partial\mathbf{t}}(\mathbf{P}_{k,t})=&\begin{pmatrix}
1/P_{k,t_2} &0  &-P_{k,t_{0}}/P^{2}_{k,t_2} \\
  0&1/P_{k,t_2}  &-P_{k,t_1}/P^{2}_{k,t_2} \\
P_{k,t_0}/l^{'}  &P_{k,t_1}/l^{'}  &P_{k,t_2}/l^{'}
\end{pmatrix},\end{aligned}
\end{equation} where 
\begin{equation}\begin{aligned}
l^{'}=&\left \| (P_{k,t_0},P_{k,t_1},P_{k,t_2})^{T}\right \|,\\
P_{k,t}=&P_{3D_{k}}+(dt*\mathcal{V}_{\text{view}_{k}})\\
=&\begin{pmatrix}
P_{k,0}+dt*v_{k,0}\\P_{k,1}+dt*v_{k,1}\\P_{k,2}+dt*v_{k,2}\end{pmatrix},\\
dt=&t_0-t,
\end{aligned}
\end{equation}
$P_{k,t}$ represents the mean of the $k^{th}$ Gaussian in ray space at timestamp $t$. And the covariance matrix $\Sigma^{'}$ in ray space is given as follows:
\begin{equation}
    \Sigma^{'}=JW_\text{view}\Sigma W_\text{view}^{T}J^{T}.
\end{equation}
Then we utilize the fast rasterization technique \cite{kerbl3Dgaussians} to get the image in the end. An algorithm outlining the process is provided as algorithm \ref{algo:project_splat}.

\begin{algorithm} 
    \textbf{Inputs:} scales $\mathbf{s}_{4D}=\{\mathbf{s}_{x},\mathbf{s}_{y},\mathbf{s}_{z},\mathbf{s}_{t}\}$,\\
  mean $\mu_{4D}=\{\mu_{x},\mu_{y},\mu_{z},\mu_{t}\}$, 
  quaternion $q=\{q_{1},q_{2},q_{3},q_{4}\}$,\\  
  velocity of mean $\mathcal{V}_{3D}=\{v_{x},v_{y},v_{z}\}$,\\
  camera position $\mathbf{P}_{camera}$,
camera rotation $\mathbf{R}_{camera}$,\\
timestamp $t_{0}$\\
  \textbf{Outputs:} rendered image  
  $I_{output}\in R^{H\times W\times 3}$\\  
  \textbf{Initialization:}\\ 
  $\mathbf{P}_{camera0}\gets None$,
  $\mathbf{R}_{camera0}\gets None$\\
  \While{Inputs != None}{
  \If{$\mathbf{P}_{camera}$ != $\mathbf{P}_{camera0}$ and $\mathbf{R}_{camera}$ != $\mathbf{R}_{camera0}$}{
    $\mathbf{S}_{3D}\gets diag(\mathbf{s}_{x},\mathbf{s}_{y},\mathbf{s}_{z})$,\\
  $\mathbf{R}_{3D}\gets q$,\\
  $\Sigma_{3D} \gets \mathbf{R}_{3D}\mathbf{S}_{3D}\mathbf{S}_{3D}^{T}\mathbf{R}_{3D}^{T}$,\\
  \Comment{calculate 3D covariance matrix $\Sigma_{3D}$}\\
  $\mathbf{P}_{view} \gets project(\mathbf{P}_{camera}$,$\mathbf{P}_{camera},\mu_{3d})$\\ \Comment{calculate mean in ray space}\\
  $\mathcal{V}_{view}\gets project(\mathbf{P}_{camera},\mathbf{P}_{camera},\mu_{3d})$\\
  \Comment{calculate velocity mean in ray space}\\
  }
  $\mathbf{P}_{t_{0}}\gets\mu_{3d}+\mathcal{V}_{3D}\times(\mu_{t}-t_{0})$\\\Comment{calculate mean in camera space at $t_{0}$}\\
  $\mathbf{P}_{view,t_{0}}\gets\mathbf{P}_{view}+\mathcal{V}_{view}\times(\mu_{t}-t_{0})$\\\Comment{calculate mean in ray space at $t_{0}$}\\
  $\Sigma^{'}\gets project(\Sigma_{3D},P_{t_{0}})$\\\Comment{approximate projection of covariance matrix}\\
    $I_{output}\gets fast\_rasterization(\mathbf{P}_{view,t_{0}},\Sigma^{'})$\\
    output $I_{output}$\\
    }
  \caption{Differentiable "projection-first" splatting for 4D Gaussian.}
  \label{algo:project_splat}
\end{algorithm}

\begin{table}[t]
  \centering
  \caption{Comparison with the state-of-the-art methods on the Plenoptic Video benchmark. *: Only tested on the scene \textit{Flame Salmon}. **: Results on \textit{Spinach},\textit{Beef},and \textit{Steak} scenes. \colorbox{red!40}{Red} denotes SOTA. \colorbox{blue!20}{Blue} denotes second. \colorbox{yellow!30}{Yellow} denotes third.}
  \begin{tabular}{@{}l|cccc}
    \toprule
    Method & PSNR$\uparrow$& DSSIM$\downarrow$& LPIPS$\downarrow$&FPS$\uparrow$\\
    \midrule
    Neural Volumes*\cite{Lombardi2019NeuralV}& 22.80& 0.062& 0.295&-\\
    LLFF*\cite{Mildenhall2019LocalLF}& 23.24& 0.076& 0.235&-\\
    
DyNeRF*\cite{9878989}& 29.58& 0.020& 0.099&0.015\\
 HexPlane\cite{Cao2023HexPlaneAF}& {{31.70}}& \colorbox{yellow!30}{0.014}& \colorbox{red!40}{{0.075}}&0.056\\
 K-Planes-explicit\cite{FridovichKeil2023KPlanesER}& 30.88& 0.020& -&0.23\\
 K-Planes-hybrid\cite{FridovichKeil2023KPlanesER}& {31.63}& \colorbox{blue!20}{{0.018}}& -&-\\
 MixVoxels-L\cite{Wang2022MixedNV}& 30.80& 0.020& 0.126&16.7\\
 StreamRF*\cite{Li2022StreamingRF}& 29.58& -& -&8.3\\
 NeRFPlayer\cite{Song2022NeRFPlayerAS}& 30.69& 0.035& 0.111&0.045\\
 HyperReel\cite{Attal2023HyperReelH6}& 31.10& 0.037& \colorbox{blue!20}{{0.096}}&2.00\\
 Deformable4DGS**\cite{Wu20234DGS}& 31.02& 0.030& 0.150&36\\
 RealTime4DGS\cite{Yang2023RealtimePD}& \colorbox{blue!20}{{32.01}}& \colorbox{yellow!30}{{0.014}}& 0.16&\colorbox{blue!20}{{72.80}}\\
 Rotor4DGS\cite{Duan20244DRotorGS}& 31.62& -& 0.14&\colorbox{yellow!30}{{277.47}}\\
 DASH\cite{Chen2025DASH4H}& \colorbox{yellow!30}{32.22}& 0.031& -&-\\
\midrule
 Ours & \colorbox{red!40}{\textbf{32.75}} & \colorbox{red!40}{\textbf{0.011}} & \colorbox{yellow!30}{\textbf{0.095}} & \colorbox{red!40}{\textbf{342.7}} \\
\bottomrule
  \end{tabular}
  \label{Tab 1.}
\end{table}

\begin{table}[ht]
  \centering
  \caption{Evaluation on Google Immersive Dataset. “Size/Fr” stands for model size per frame. \colorbox{red!40}{Red} denotes SOTA. \colorbox{blue!20}{Blue} denotes second. \colorbox{yellow!30}{Yellow} denotes third.}
  \begin{tabular}{@{}l|ccccc}
    \toprule
    Method & PSNR$\uparrow$ & LPIPS$\downarrow$ & DSSIM$\downarrow$&  Size/Fr$\downarrow$ & FPS$\uparrow$ \\
    \midrule
     NeRFPlayer\cite{Pumarola2020DNeRFNR} & 25.8 & 0.196 & 0.076 & \colorbox{blue!20}{17.1 MB} & 0.12 \\
     HyperReel\cite{Pumarola2020DNeRFNR} & \colorbox{blue!20}{28.8} & \colorbox{blue!20}{0.193} & \colorbox{blue!20}{0.063} & \colorbox{yellow!30}{1.2 MB} & \colorbox{blue!20}{4} \\
    SpacetimeGS\cite{Li2023SpacetimeGF} & \colorbox{yellow!30}{29.2} & \colorbox{red!40}{\textbf{0.081}}& \colorbox{yellow!30}{0.042} & \colorbox{yellow!30}{1.2 MB} & \colorbox{yellow!30}{99} \\
    \midrule
Ours & {\colorbox{red!40}{\textbf{30.6}}} & \colorbox{yellow!30}{0.104}& \colorbox{red!40}{\textbf{0.041}} & \colorbox{red!40}{\textbf{0.95 MB}} & \colorbox{red!40}{\textbf{183}}\\
    \bottomrule
  \end{tabular}
  \label{Tab 2.}
\end{table}
 \subsection{Flow-Gradient Guided Consistency Loss and Temporal Splitting Strategy}
 Beyond optimizing rendering efficiency, we also aim to enhance rendering quality and ensure coherent object motions in dynamic scenes. Artifacts and distortions, which already degrade 3DGS rendering quality \cite{Yu2023MipSplattingA3}, are even more problematic in dynamic settings where inaccurate motion boundaries or inconsistent trajectories may occur. To tackle these challenges, we introduce a flow-gradient guided consistency loss and a temporal splitting strategy.
 \subsubsection{Flow-Gradient Guided Consistency Loss}
Most existing dynamic scene methods only supervise the rendered appearance, without explicitly constraining the motion of objects. This often leads to distorted or unstable trajectories, especially around motion boundaries. Since both the color image and optical flow are rendered outputs in our framework, they can be easily obtained without requiring additional annotations. We introduce a regularization to enforce structural consistency between them. 

Concretely, we regard image gradients as reliable indicators of structural discontinuities. We first compute the gradient magnitude of the rendered optical flow field:
\begin{equation}
M=\sqrt{u^2+v^2+\varepsilon},
\end{equation}
and extract the gradient of the RGB image $I$ using a Sobel operator. We then design a consistency loss to encourage flow discontinuities to align with image edges:
\begin{equation}
L_\text{fg} = \lambda_{flow} \times( \frac{1}{N}\sum_{x,y} \|\nabla M(x,y)\| \cdot \big(1 - \|\nabla I(x,y)\|\big)),
\end{equation}
where $\nabla M(x,y)$ denotes the normalized gradient of the flow magnitude, $\nabla I(x,y)$ is the normalized image gradient and $\lambda_{flow}$ is the scale of flow-gradient guided consistency loss. 

This loss penalizes strong flow gradients in regions without corresponding image edges, while allowing sharp motion changes at true structural boundaries. By doing so, the flow-gradient guided consistency loss improves the coherence of rendered motion and produces more faithful dynamic trajectories, thereby enhancing the realism of novel view synthesis in dynamic scenes.
\subsubsection{Temporal Splitting Strategy}
In addition to spatial constraints, dynamic scenes often involve complex, non-linear object motions such as sudden appearance, disappearance, or occlusion changes. Conventional 4DGS approaches approximate object motions with first- or second-order models, which is insufficient for capturing such dynamics. To address this, we decouple the splitting strategy into temporal and spatial components. Specifically, time-domain splitting is guided by the gradient of the $t$ coordinate, while spatial splitting is based on the gradients of $xyz$ coordinates. Moreover, the spatial and temporal scales of 4D Gaussians are split independently, allowing finer-grained modeling of both spatial structures and temporal variations.
\section{Experiments}
\subsection{Datasets}

\begin{figure*}
 \centering
    \begin{subfigure}{\linewidth}
    \hspace{1cm}
        \includegraphics[width=0.9\linewidth,height=0.6\linewidth]{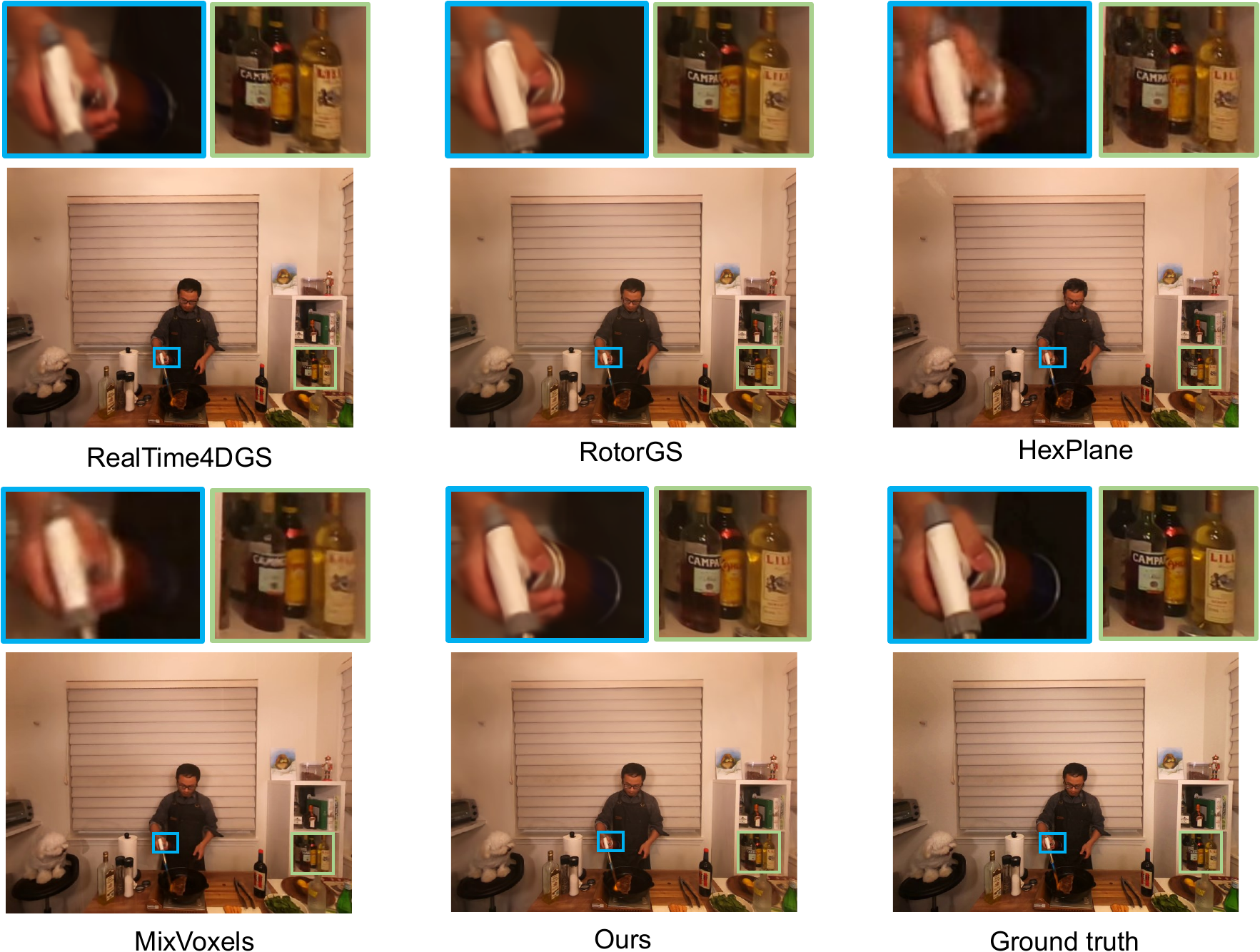}
    \caption{Comparison with the state-of-the-art on \textit{flame steak} scene}
    \label{}
    \end{subfigure}

\begin{subfigure}{\linewidth}
\hspace{1cm}
        \includegraphics[width=0.9\linewidth,height=0.6\linewidth]{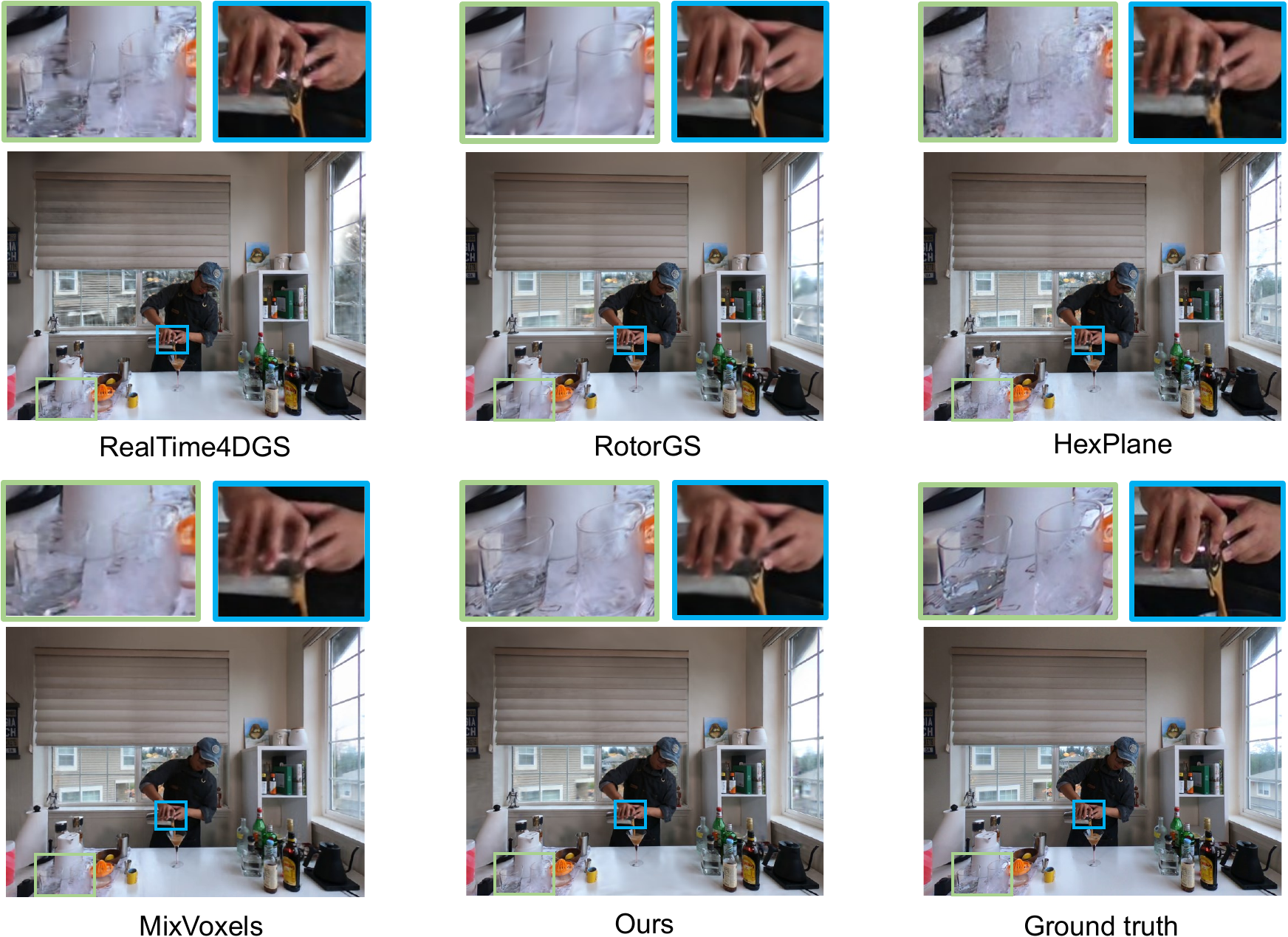}
    \caption{Comparison with the state-of-the-art approaches on \textit{coffee martini} scene}
    \label{}
    \end{subfigure}
    \caption{Visual comparisons on Plenoptic Video Dataset}
    \label{fig:dynerf}
\end{figure*}

\begin{figure*}
   \begin{subfigure}{0.3\linewidth}
        \includegraphics[width=1\linewidth,height=0.6\linewidth]{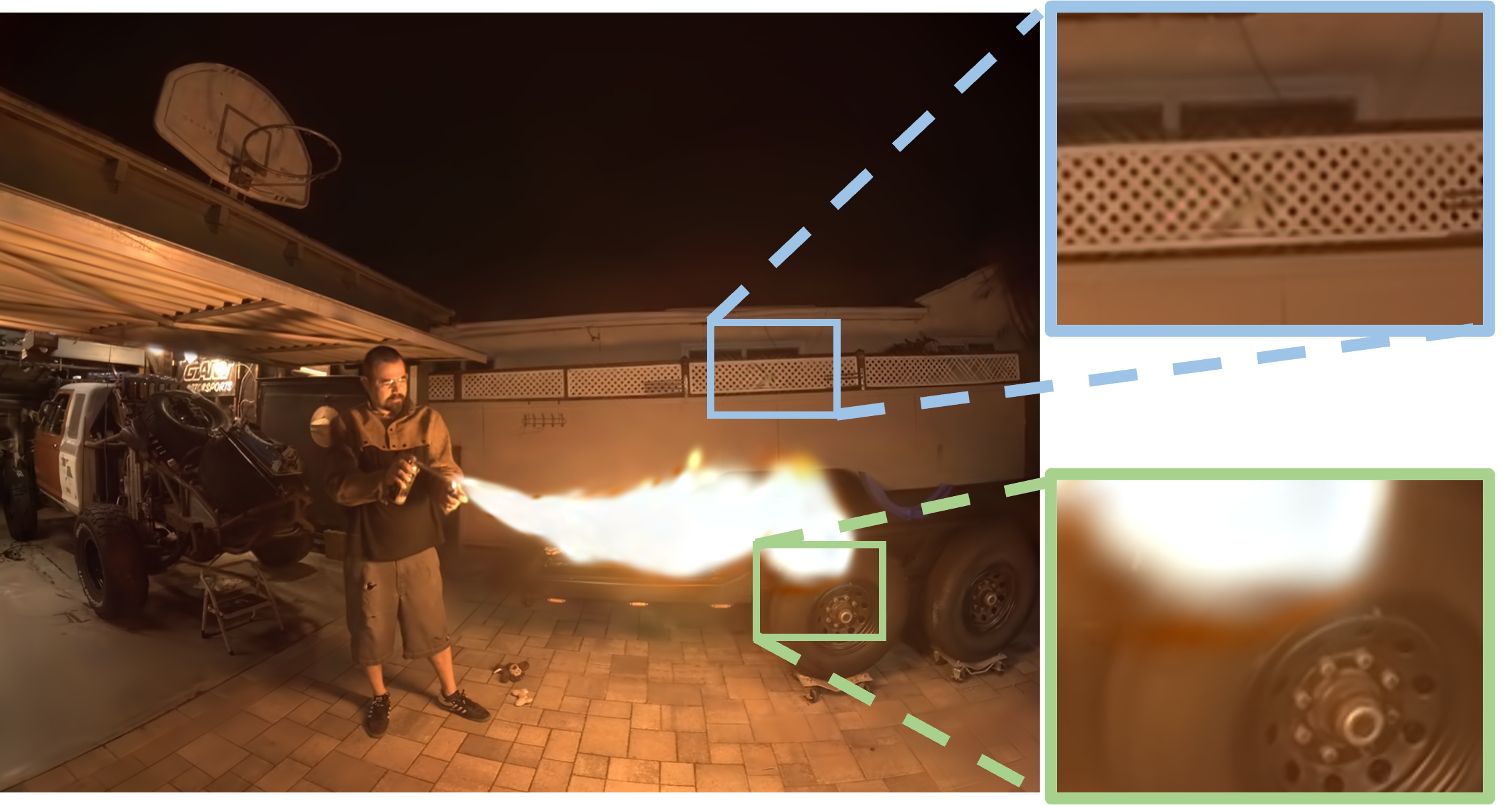}
    \caption{SpacetimeGS\cite{Li2023SpacetimeGF}}
    \label{}
    \end{subfigure}
    \centering
    \begin{subfigure}{0.3\linewidth}
        \includegraphics[width=1\linewidth,height=0.6\linewidth]{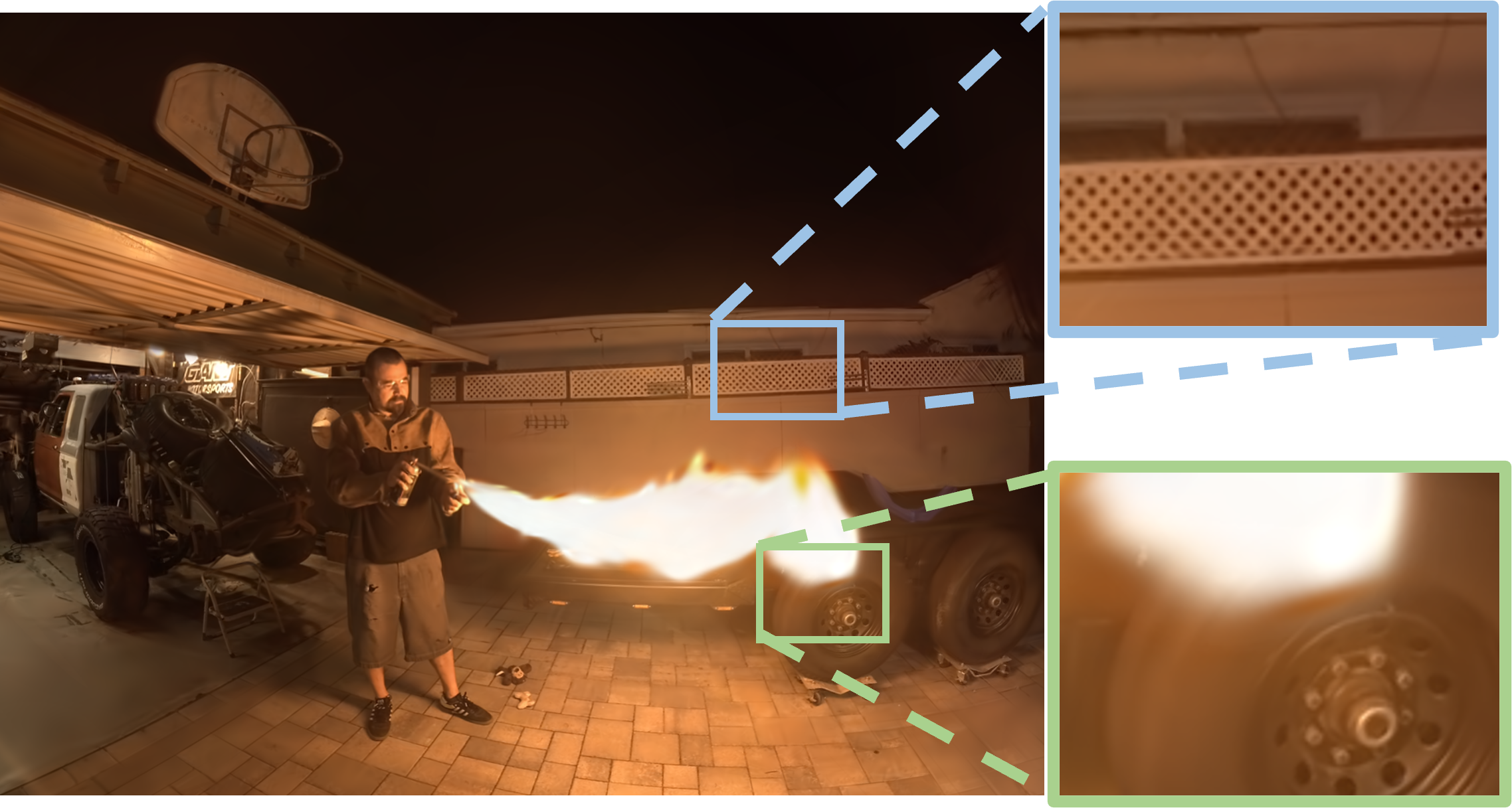}
    \caption{Ours}
    \label{}
    \end{subfigure}
    \centering
    \begin{subfigure}{0.3\linewidth}
        \includegraphics[width=1\linewidth,height=0.6\linewidth]{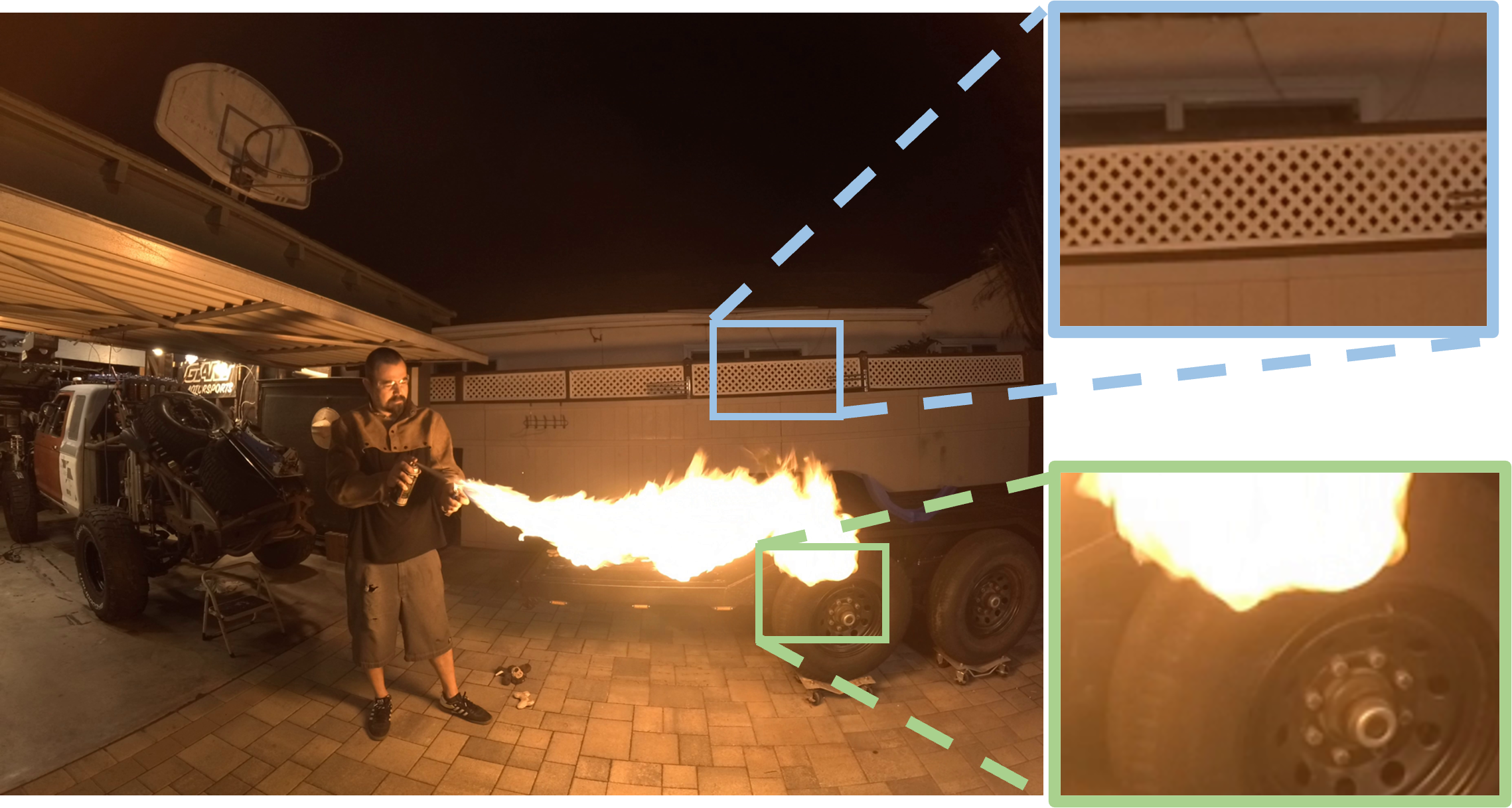}
    \caption{Ground Truth}
    \label{}
    \end{subfigure}
    \centering
    \caption{Visual comparisons on Google Immersive Dataset}
    \label{fig:immersive}
\end{figure*}

\begin{table}[t]
  \centering
  \caption{Evaluation on HyperNeRF Dataset. \colorbox{red!40}{Red} denotes SOTA. \colorbox{blue!20}{Blue} denotes second. \colorbox{yellow!30}{Yellow} denotes third.}
  \begin{tabular}{@{}c|cccc|c}
    \toprule
    Method & Chicken & Banana & Broom & 3D Printer & Avg \\
    \midrule
Nerfies\cite{Pumarola2020DNeRFNR} & 26.7 & 22.4 & 19.2 & 20.6&22.2\\
    HyperNeRF\cite{Pumarola2020DNeRFNR} & 26.9 & 23.3& 19.3& 20.0& 22.4\\
    TiNeuVox-B\cite{Fang2022FastDR} & \colorbox{blue!20}{28.3}  & \colorbox{blue!20}{24.4} & 21.5& \colorbox{red!40}{\textbf{22.8}}& \colorbox{blue!20}{24.3}\\
    FFDNeRF\cite{Cao2023HexPlaneAF} & 28.0  & 24.3 & \colorbox{blue!20}{21.9}& \colorbox{red!40}{\textbf{22.8}}&24.2\\
    3D-GS\cite{FridovichKeil2023KPlanesER} & 19.7  & 20.4 & 20.6& 18.3&19.7\\
    Deformable4DGS\cite{FridovichKeil2023KPlanesER} & \colorbox{yellow!30}{28.7}  & \colorbox{yellow!30}{28.0} & \colorbox{yellow!30}{22.0}& \colorbox{blue!20}{22.1}&\colorbox{yellow!30}{25.2}\\
    \midrule
    Ours & \colorbox{red!40}{\textbf{29.4}}& \colorbox{red!40}{\textbf{29.1}}& \colorbox{red!40}{\textbf{22.4}}& \colorbox{yellow!30}{22.3}&\colorbox{red!40}{\textbf{25.8}}\\
    \bottomrule
  \end{tabular}
  \label{Tab 3.}
\end{table}

\begin{table}[t]
  \centering
  \caption{Evaluation on D-NeRF Dataset. \colorbox{red!40}{Red} denotes SOTA. \colorbox{blue!20}{Blue} denotes second. \colorbox{yellow!30}{Yellow} denotes third.}
  \begin{tabular}{@{}l|ccccc}
    \toprule
    Method & PSNR$\uparrow$ & SSIM$\uparrow$ & LPIPS$\downarrow$ & FPS$\uparrow$ \\
    \midrule
    \multicolumn{5}{l}{-D-NeRF (synthetic, monocular)} \\
    \midrule
    T-NeRF\cite{Pumarola2020DNeRFNR} & 29.51 & \colorbox{blue!20}{0.95} & 0.08 & - \\
    D-NeRF\cite{Pumarola2020DNeRFNR} & 29.67 & \colorbox{blue!20}{0.95} & 0.07 & - \\
    TiNeuVox\cite{Fang2022FastDR} & 32.67  & \colorbox{yellow!30}{0.97} & \colorbox{blue!20}{0.04} & 1.60 \\
    HexPlanes\cite{Cao2023HexPlaneAF} & 31.04  & \colorbox{yellow!30}{0.97} & - & - \\
    K-Planes-explicit\cite{FridovichKeil2023KPlanesER} & 31.05  & \colorbox{yellow!30}{0.97} & - & - \\
    K-Planes-hybrid\cite{FridovichKeil2023KPlanesER} & 31.61  & \colorbox{yellow!30}{0.97} & \colorbox{red!40}{\textbf{0.02}} & - \\
    V4D\cite{Gan2022V4DVF} & \colorbox{yellow!30}{33.72} & \colorbox{red!40}{\textbf{0.98}} & \colorbox{yellow!30}{0.03} & 2.08 \\
    RealTime4DGS\cite{Yang2023RealtimePD} & 32.71  & \colorbox{yellow!30}{0.97} & \colorbox{yellow!30}{0.03} & \colorbox{blue!20}{{289.07}} \\
    Rotor4DGS\cite{Duan20244DRotorGS} & {\colorbox{red!40}{\textbf{34.26}}}  & \colorbox{yellow!30}{0.97} & \colorbox{yellow!30}{0.03} & \colorbox{yellow!30}{{1257.63}} \\
    \midrule
    Ours & \colorbox{blue!20}{33.61}  & \colorbox{red!40}{\textbf{0.98}} & \colorbox{red!40}{\textbf{0.02}} & \colorbox{red!40}{\textbf{1549.03}} \\
    \bottomrule
  \end{tabular}
  \label{Tab 4.}
\end{table}

\begin{figure*}
    \centering
    \begin{subfigure}{0.23\linewidth}
        \includegraphics[width=0.9\linewidth,height=2\linewidth]{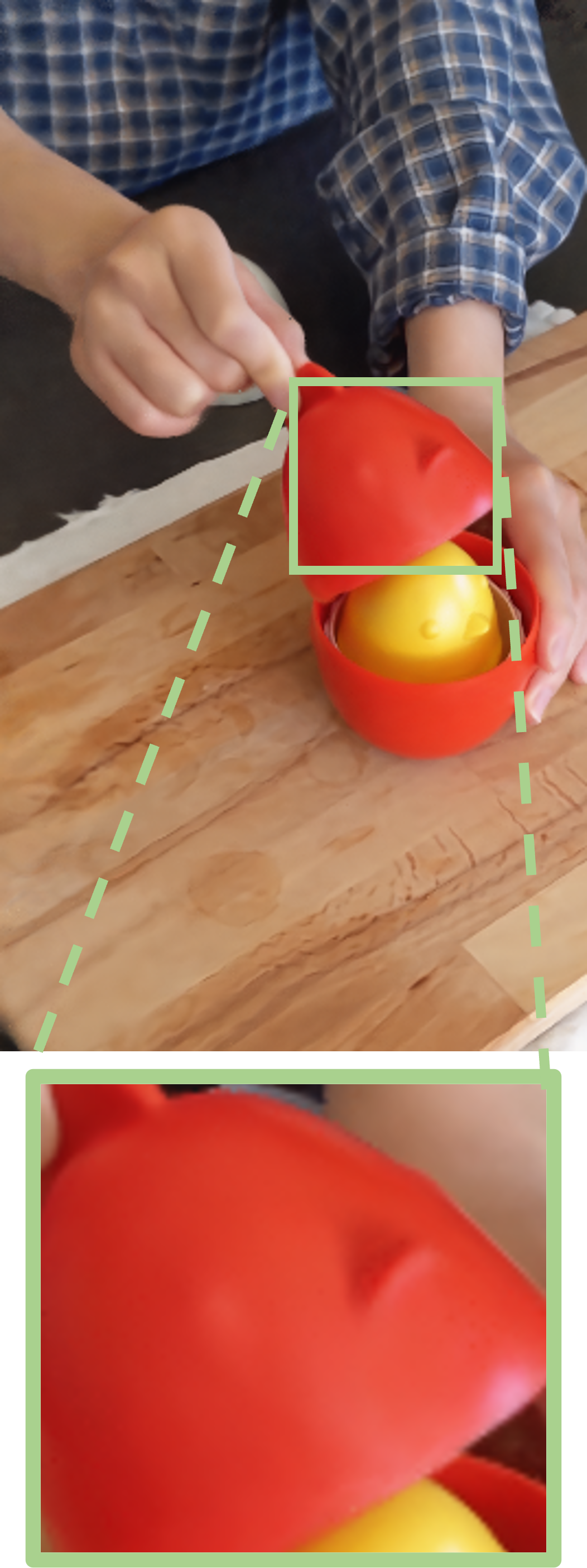}
    \caption{RealTime4DGS\cite{Yang2023RealtimePD}}
    \label{}
    \end{subfigure}
    \centering
    \begin{subfigure}{0.23\linewidth}
        \includegraphics[width=0.9\linewidth,height=2\linewidth]{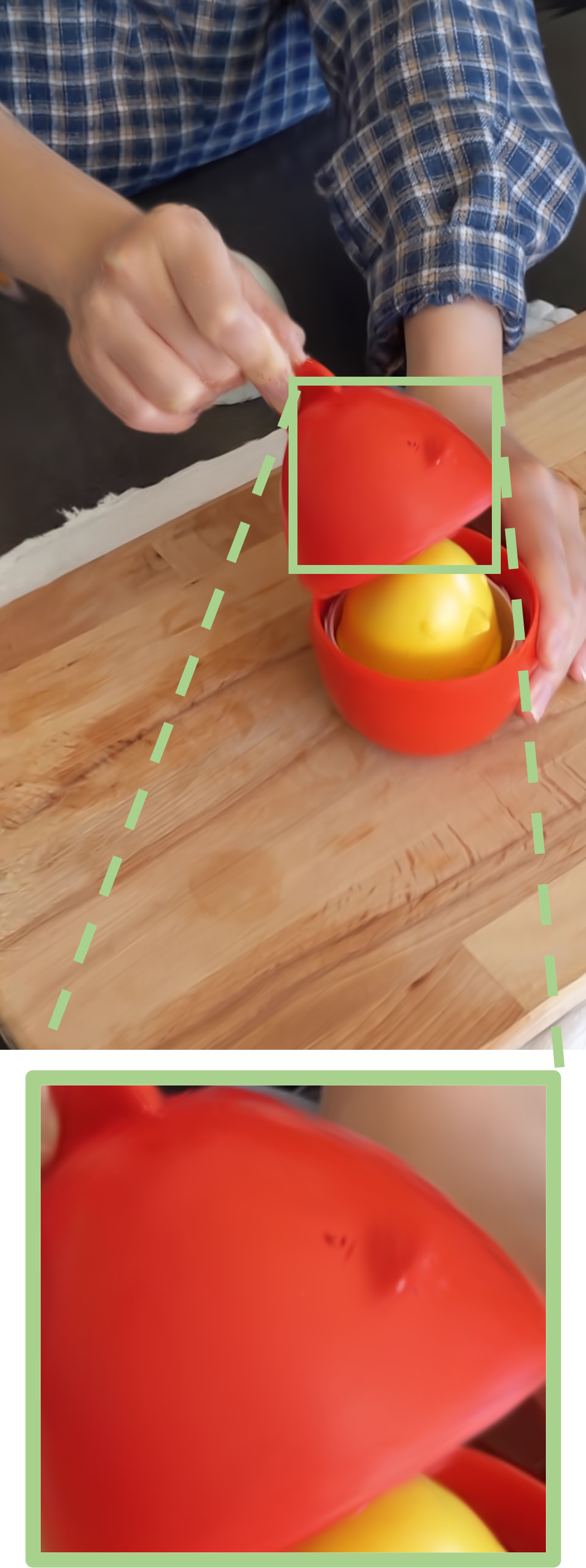}
    \caption{Rotor4DGS\cite{Duan20244DRotorGS}}
    \label{}
    \end{subfigure}
    \centering
    \begin{subfigure}{0.23\linewidth}
        \includegraphics[width=0.9\linewidth,height=2\linewidth]{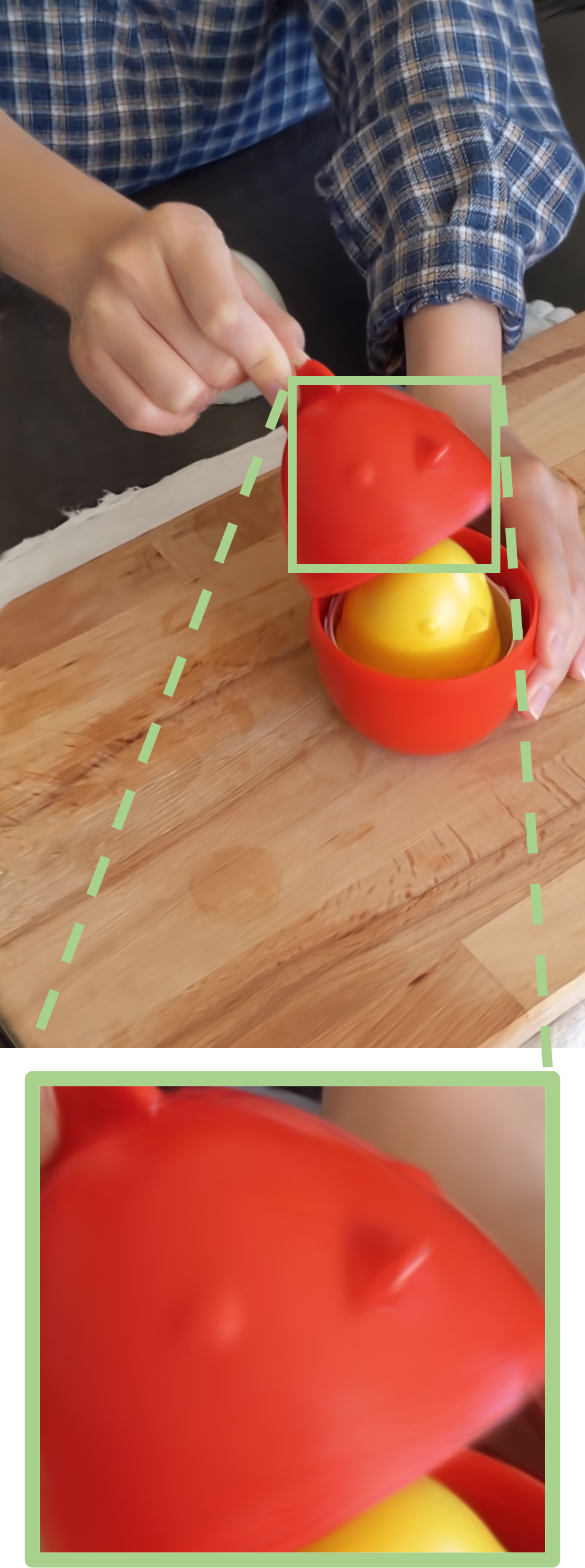}
    \caption{Ours}
    \label{}
    \end{subfigure}
    \centering
    \begin{subfigure}{0.23\linewidth}
        \includegraphics[width=0.9\linewidth,height=2\linewidth]{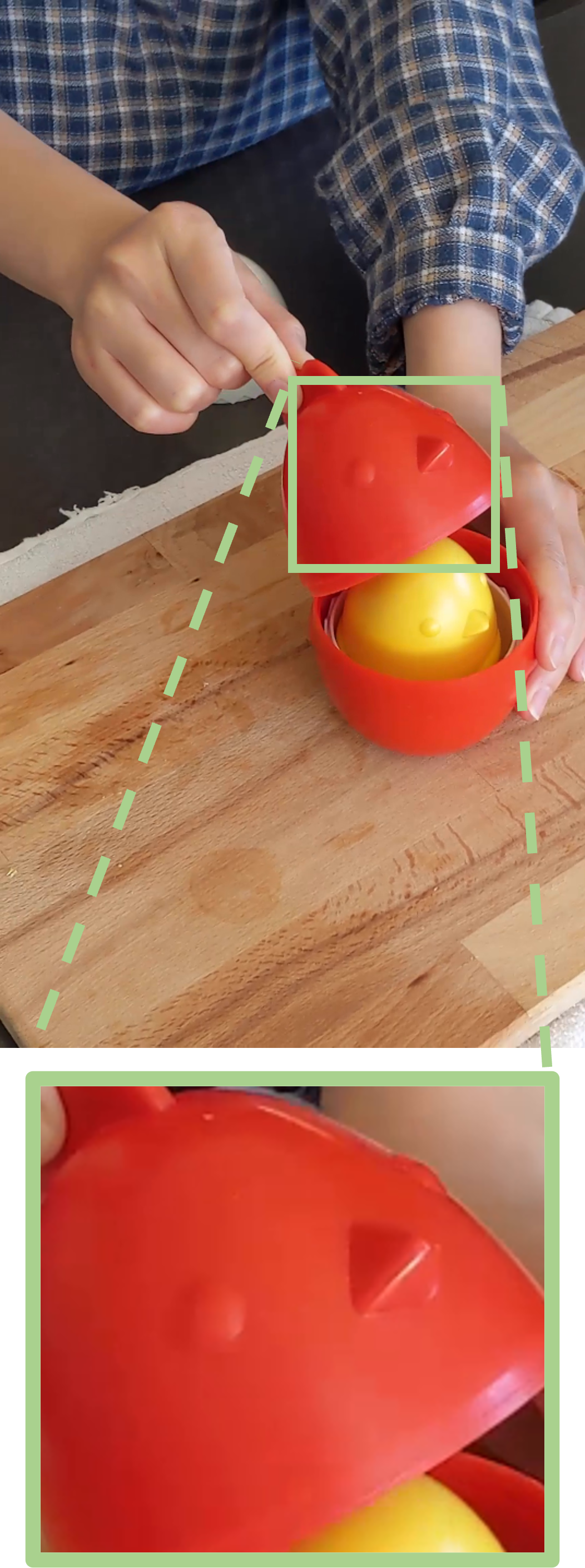}
    \caption{Ground Truth }
    \end{subfigure}
    \caption{Visualization of \textit{Chicken} scene on HyperNeRF Dataset}
    \label{fig:chicken}
\end{figure*}

\begin{figure*}
    \centering
    \begin{subfigure}{0.23\linewidth}
        \includegraphics[width=0.9\linewidth,height=2\linewidth]{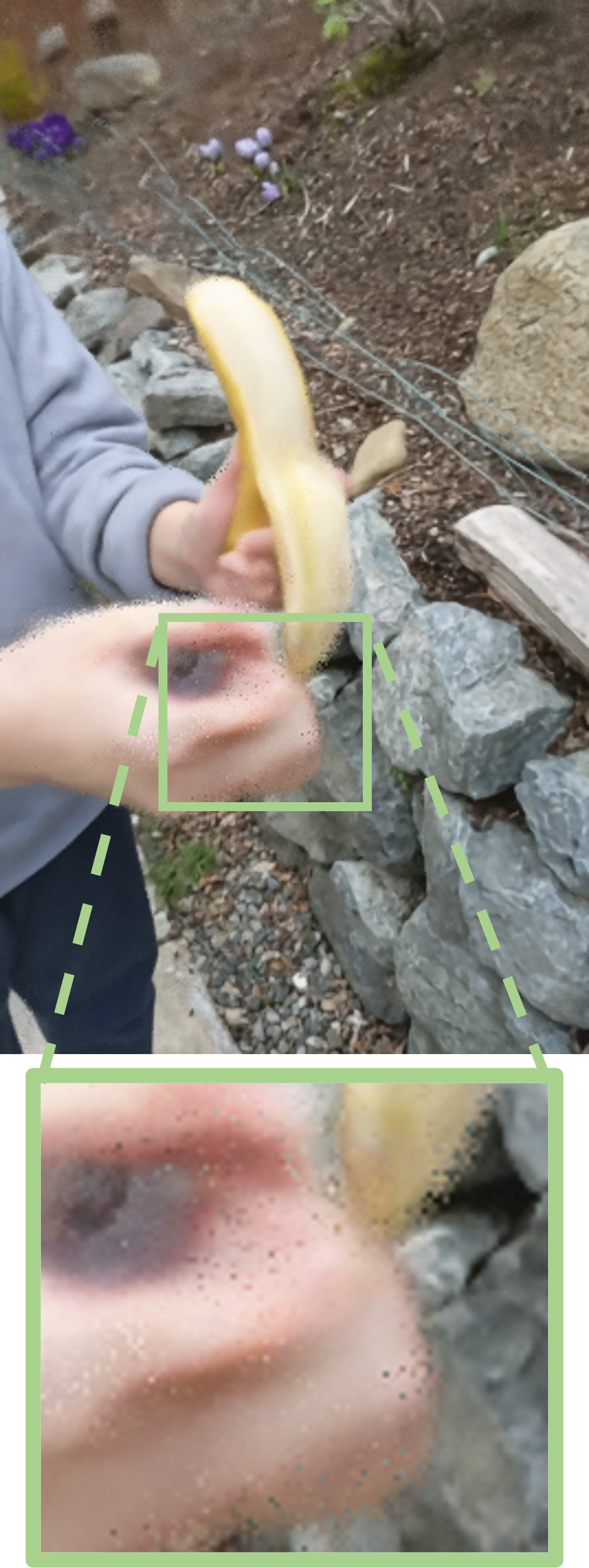}
    \caption{RealTime4DGS\cite{Yang2023RealtimePD}}
    \label{}
    \end{subfigure}
    \centering
    \begin{subfigure}{0.23\linewidth}
        \includegraphics[width=0.9\linewidth,height=2\linewidth]{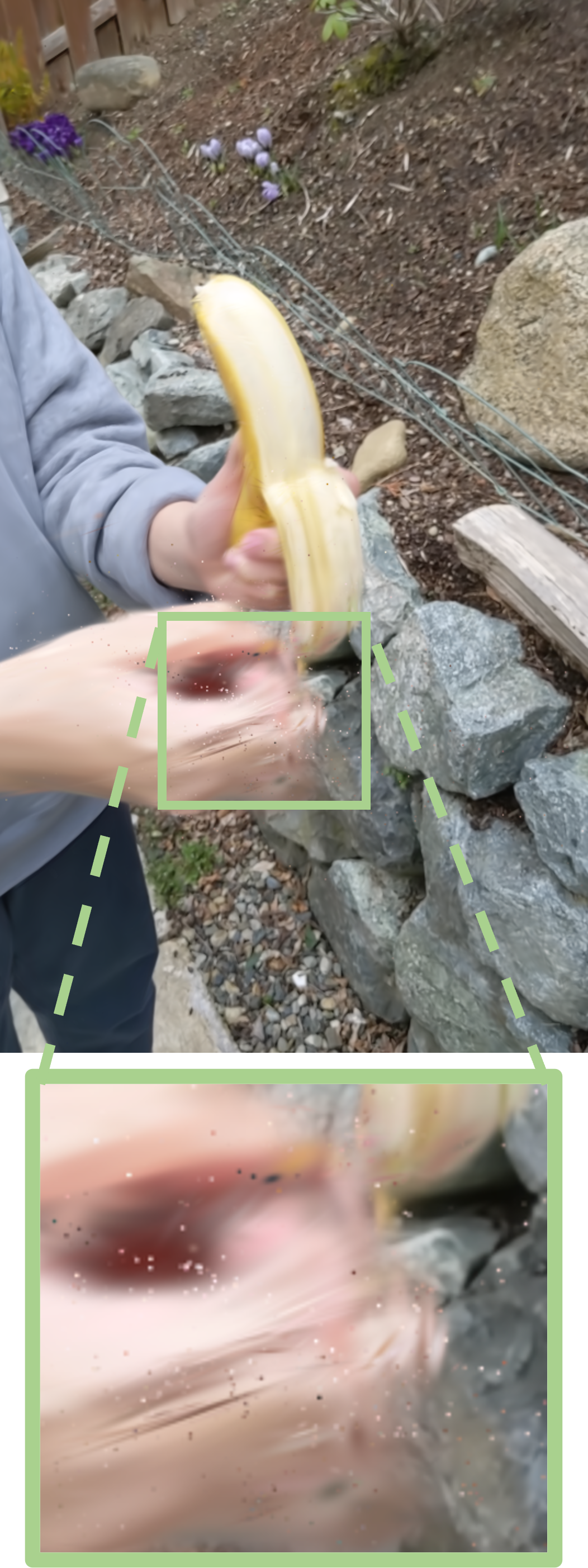}
    \caption{Rotor4DGS\cite{Duan20244DRotorGS}}
    \label{}
    \end{subfigure}
    \centering
    \begin{subfigure}{0.23\linewidth}
        \includegraphics[width=0.9\linewidth,height=2\linewidth]{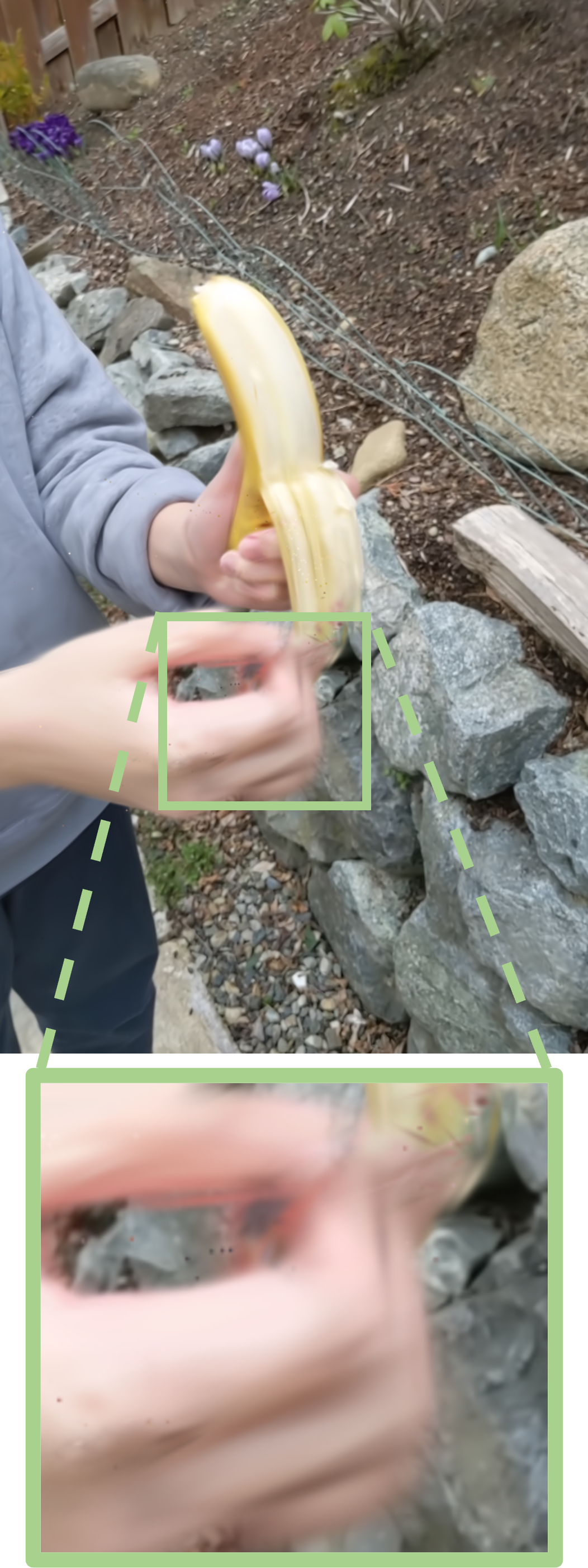}
    \caption{Ours}
    \label{}
    \end{subfigure}
    \centering
    \begin{subfigure}{0.23\linewidth}
        \includegraphics[width=0.9\linewidth,height=2\linewidth]{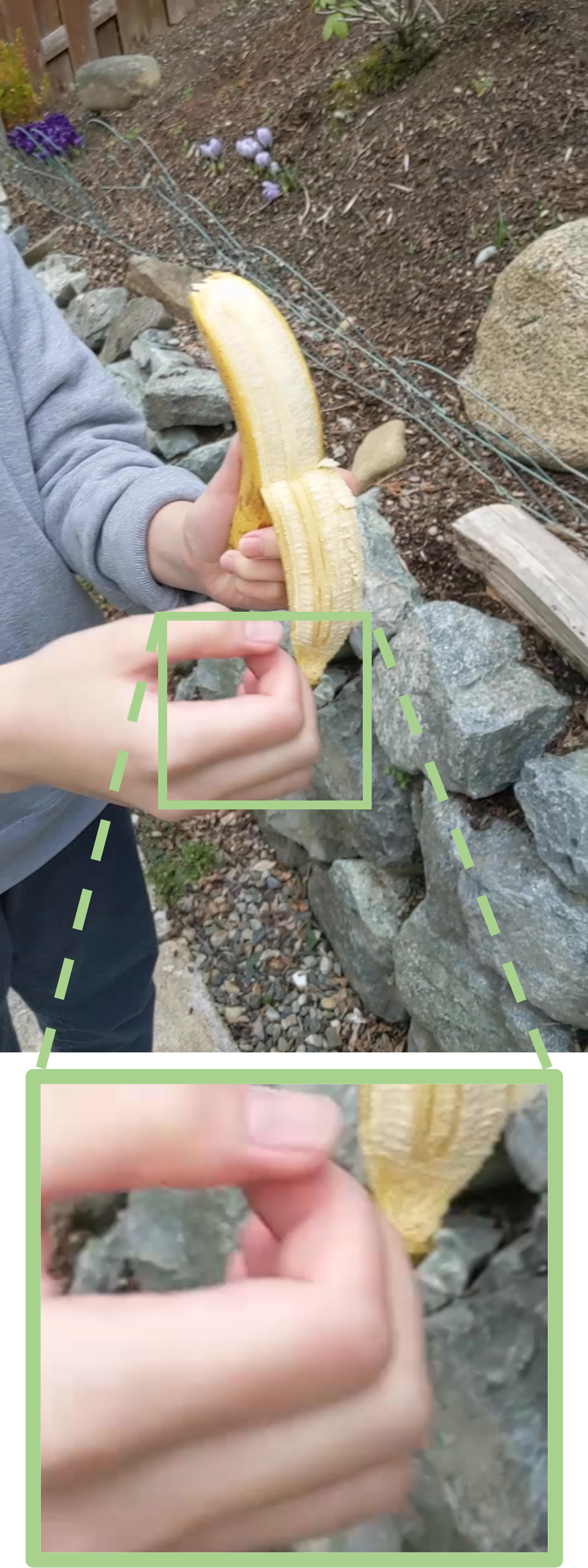}
    \caption{Ground Truth }
    \end{subfigure}
    \caption{Visualization of \textit{Banana} scene on HyperNeRF Dataset}
    \label{fig:banana}
\end{figure*}

    \begin{figure*}
    \centering
    \begin{subfigure}{0.23\linewidth}
        \includegraphics[width=0.9\linewidth,height=0.8\linewidth]{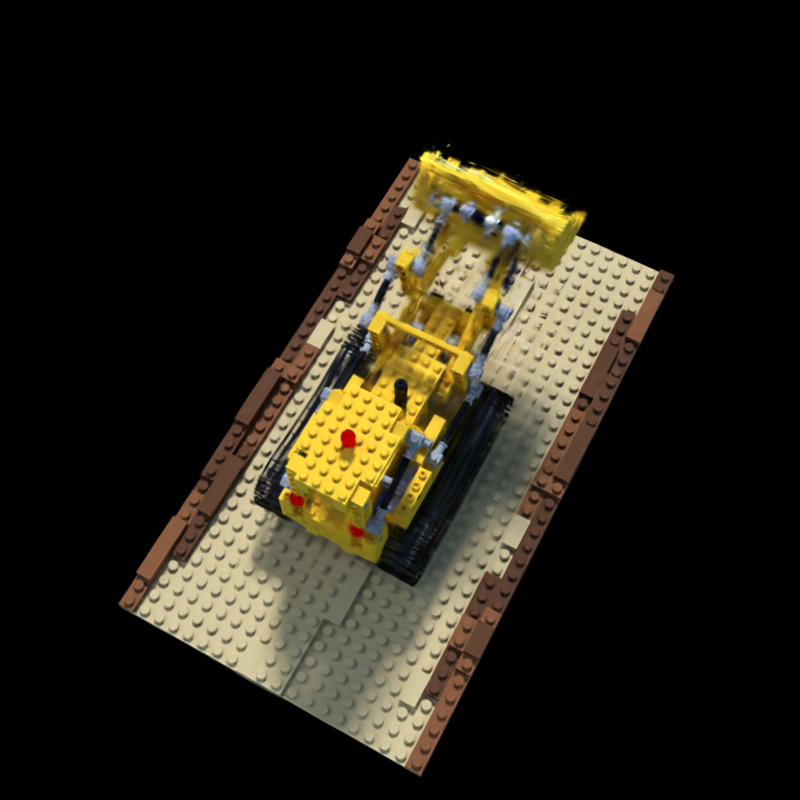}
    \caption{RealTime4DGS\cite{Yang2023RealtimePD}}
    \label{}
    \end{subfigure}
    \centering
    \begin{subfigure}{0.23\linewidth}
        \includegraphics[width=0.9\linewidth,height=0.8\linewidth]{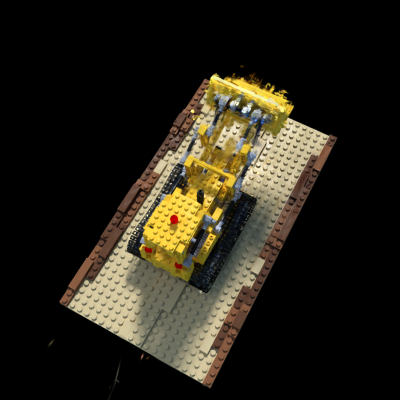}
    \caption{Rotor4DGS\cite{Duan20244DRotorGS}}
    \label{}
    \end{subfigure}
    \centering
    \begin{subfigure}{0.23\linewidth}
        \includegraphics[width=0.9\linewidth,height=0.8\linewidth]{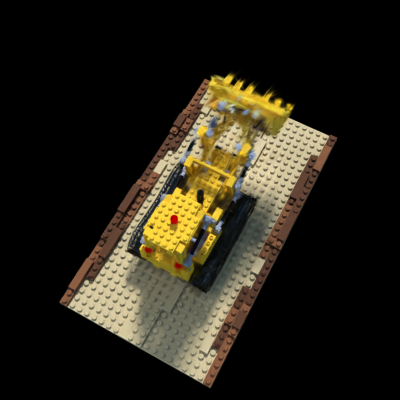}
    \caption{Ours}
    \label{}
    \end{subfigure}
    \centering
    \begin{subfigure}{0.23\linewidth}
        \includegraphics[width=0.9\linewidth,height=0.8\linewidth]{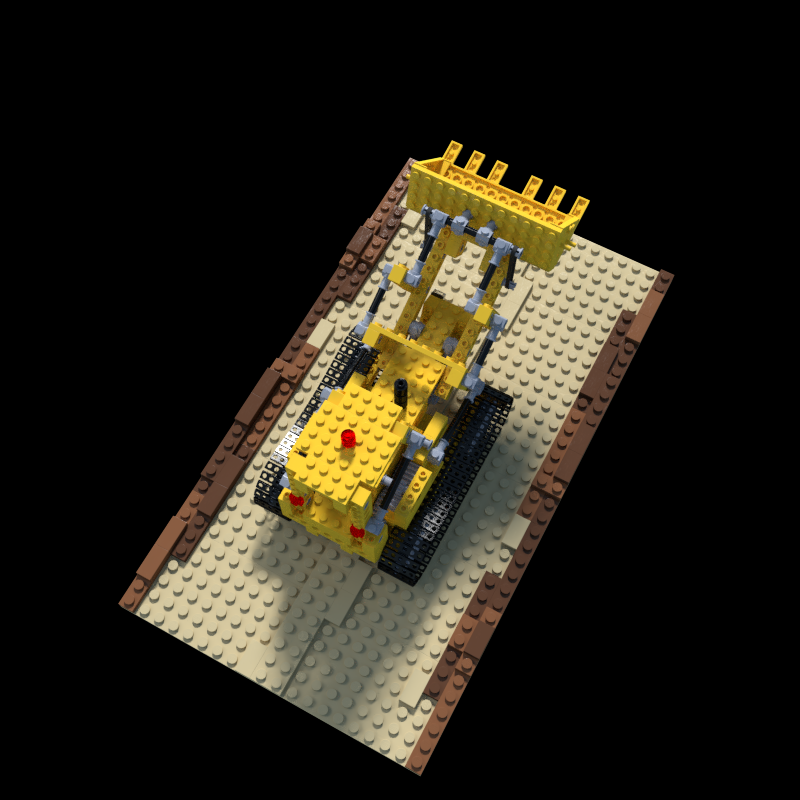}
    \caption{Ground Truth }
    \end{subfigure}
    \caption{Visualization of \textit{Lego} scene}
    \label{fig:lego}
\end{figure*}

In this section, we evaluate our method on four commonly used datasets, Plenoptic Video Dataset \cite{9878989}, Google Immersive Dataset\cite{10.1145/3386569.3392485}, HyperNeRF Dataset\cite{Park2021HyperNeRF} and D-NeRF dataset \cite{Pumarola2020DNeRFNR}. 

\textbf{Plenoptic Video Dataset}\cite{9878989} contains real-world multiview videos of 6 scenes, each lasting ten seconds. These scenes include complex motions and materials that are reflective or transparent.  Following prior work \cite{Yang2023RealtimePD}, we utilize the colored point cloud generated by COLMAP from the first frame of each scene. And the resolution of \begin{math}
    1352\times1014
\end{math} is adopted.

\textbf{Google Immersive Dataset}\cite{10.1145/3386569.3392485} contains both indoor and outdoor scenes captured with a $46$-camera rig. The cameras are equipped with fisheye lenses and mounted on an outward-facing hemisphere, which results in less view overlap compared to traditional outside-in setups, thereby posing additional challenges. Following prior work \cite{Attal2023HyperReelH6,Song2022NeRFPlayerAS}, we evaluate on $7$ selected scenes (Welder, Flames, Truck, Exhibit, FacePaint1, FacePaint2, Cave) and hold out the center camera as the test view.

\textbf{HyperNeRF Dataset}\cite{Park2021HyperNeRF} contains dynamic real-world scenes exhibiting complex non-rigid deformations and topological changes. The data is captured using one or two handheld cameras with relatively straightforward camera motion. Following prior work \cite{Park2021HyperNeRF}, we utilize $200$ randomly selected frames from each sequence for training and evaluation.

\textbf{D-NeRF}\cite{Pumarola2020DNeRFNR} is a monocular video dataset comprising eight videos of synthetic scenes. Following prior work \cite{Yang2023RealtimePD}, we utilize randomly selected points, evenly distributed within the cubic volume defined by \begin{math}
    [-1.2,1.2]^{3}
\end{math}, and set their initial mean as the time duration of scene.
\subsection{Implementation Details}

The densification gradient threshold is set as \begin{math}
    5e-5
\end{math} in D-NeRF and \begin{math}
    2e-5
\end{math} in Plenoptic datasets. The rotors are initialized with (1,0,0,0) equivalent of static identity transformation. Learning rates, densification, pruning, and
opacity reset settings all follow \cite{kerbl3Dgaussians}. Optimizer is Adam following prior work. For each scene we train for 20,000 steps. The LPIPS \cite{Zhang2018TheUE} in the Plenoptic Video dataset and the Google Immersive Dataset\cite{10.1145/3386569.3392485} are computed using AlexNet \cite{Krizhevsky2012ImageNetCW}. The D-NeRF \cite{Pumarola2020DNeRFNR} dataset is computed using VGGNet \cite{Simonyan2014VeryDC}, respectively. To ensure a fair comparison with previous works we do not fix the viewpoints.
\subsection{Comparison with Existing Works on Dynamic Novel View Synthesis}
\subsubsection{Results on the multi-view real-world scenes}

\textbf{Plenoptic Video Dataset} \cite{9878989}

As summarized in Tab.~\ref{Tab 1.}, our approach substantially surpasses previous methods in both rendering quality and computational efficiency on the Plenoptic Video Dataset~\cite{9878989}. Specifically, our method renders high-resolution videos (\begin{math}
    1352\times1014
\end{math}) at 343 FPS on an NVIDIA RTX 3090 GPU, representing a significant improvement over all existing approaches. Notably, even when compared with recent slicing-first 4D Gaussian methods~\cite{Wu20234DGS,Duan20244DRotorGS,Yang2023RealtimePD}, our technique exhibits superior rendering fidelity and speed. The comparison with~\cite{Yang2023RealtimePD} further highlights the efficiency and visual quality advantages of our Disentangled4DGS framework. Visual results in Fig.~\ref{fig:dynerf} demonstrate that our method preserves finer dynamic and static details—such as the spray gun in flame steak and the glass cups in coffee martini—surpassing prior methods in both realism and consistency.

\textbf{Google Immersive Dataset} \cite{10.1145/3386569.3392485}

As presented in Tab.~\ref{Tab 2.}, we evaluate PSNR, LPIPS, DSSIM, model size per frame, and FPS. Our method achieves the best overall performance across all metrics on a single NVIDIA A6000 GPU under the same experimental settings as~\cite{Li2023SpacetimeGF}: the highest PSNR (30.6), the lowest DSSIM (0.041), the most compact model size (0.95 MB/frame), and the fastest rendering speed (194 FPS). Compared with SpacetimeGS~\cite{Li2023SpacetimeGF}, our approach improves PSNR by 1.4 dB while nearly doubling the rendering speed, demonstrating an excellent trade-off among quality, compactness, and efficiency. As shown in Fig.~\ref{fig:immersive}, our method also produces fewer visual artifacts, further confirming its robustness in complex immersive scenes.

\textbf{HyperNeRF Dataset} \cite{Park2021HyperNeRF}

Following prior works, we report PSNR values on four challenging dynamic scenes (Chicken, Banana, Broom, 3D Printer) in the HyperNeRF dataset (Tab.~\ref{Tab 3.}). Our method consistently outperforms all baselines, achieving the best results on three out of four scenes and an average PSNR of 25.8, surpassing Deformable4DGS~\cite{FridovichKeil2023KPlanesER} by +0.6 dB on average. Despite comparable performance on 3D Printer, our approach exhibits stronger generalization and motion modeling capabilities, particularly for complex dynamic regions—such as the moving chicken toy in Chicken and the hand motion in Banana—as illustrated in Fig.~\ref{fig:chicken} and Fig.~\ref{fig:banana}.

\subsubsection{Results on the monocular synthetic videos} 

Monocular video novel view synthesis for dynamic scenes remains particularly challenging due to the sparsity of input views \cite{Duan20244DRotorGS}. As summarized in Tab.\ref{Tab 4.}, our method achieves the fastest rendering speed among all evaluated techniques, while also delivering a balanced level of quality. A key advantage of our approach is its consistent performance across all types of scenes. This contrasts with methods like Rotor4DGS \cite{Duan20244DRotorGS}, which can exhibit unstable performance in certain scenarios, such as the Lego scene as shown in Fig. \ref{fig:lego}. In such cases, the PSNR (24.93) is significantly lower than our method’s 26.60, demonstrating our superior stability. The variance in PSNR across scenes further confirms the robustness of our method. Detailed PSNR performance for each scene is presented in the \textbf{supplementary material}.

\subsubsection{Speed Result of Same Rendering Condition}
\begin{table}
    \caption{Rendering time comparison on 100,000 Gaussians}
    \centering
    \begin{tabular}{c|c}
    \toprule
         Method&  Time(ms)\\
         \midrule
         Rotor4DGS\cite{Duan20244DRotorGS}&  0.94\\
         \midrule
          Ours&  0.69\\
          \bottomrule
    \end{tabular}
    \label{tab:speed}
\end{table}
\begin{table}
    \centering
        \caption{Ablation studies. We validate two designs on rendering quality on D-NeRF Dataset. }
    \begin{tabular}{lcc|clll}
    \toprule
           ID&\multicolumn{2}{c|}{Ablation Items}&  \multicolumn{2}{c}{D-NeRF} & \multicolumn{2}{c}{Plenoptic Dataset}\\
           &\makecell{Edge\\Loss}&\makecell{Temporal\\Split}& 
     PSNR$\uparrow$&SSIM$\uparrow$ & PSNR$\uparrow$&DSSIM$\downarrow$\\
     \midrule
  (a)&& & 33.20&0.95 & 32.44&0.013\\
  (b)&& \checkmark& 33.47&0.97 & 32.58&0.012\\
  (c)&\checkmark& & 33.40&0.97 & 32.56&0.013\\
  \midrule
  Full&\checkmark& \checkmark& 33.61&0.98 & 32.75&0.011\\
  \bottomrule
  \end{tabular}
    \label{Tab 6.}
\end{table}

\begin{figure*}
    \centering
    \begin{subfigure}{0.23\linewidth}
        \includegraphics[width=\linewidth,height=2\linewidth]{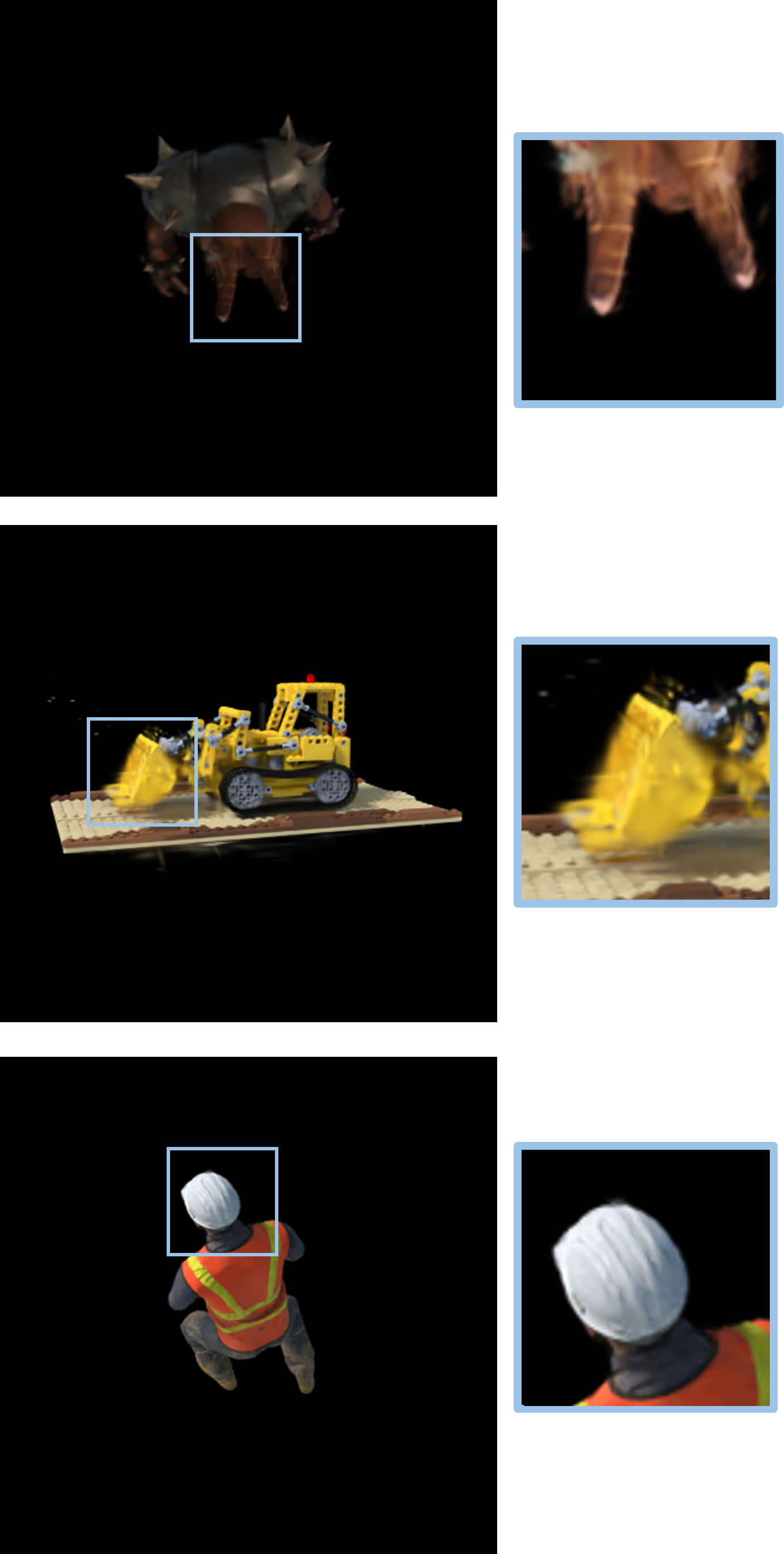}
    \caption{w/ Temporal Splitting\\ \& Edge Loss}
    \label{}
    \end{subfigure}
    \centering
    \begin{subfigure}{0.23\linewidth}
        \includegraphics[width=\linewidth,height=2\linewidth]{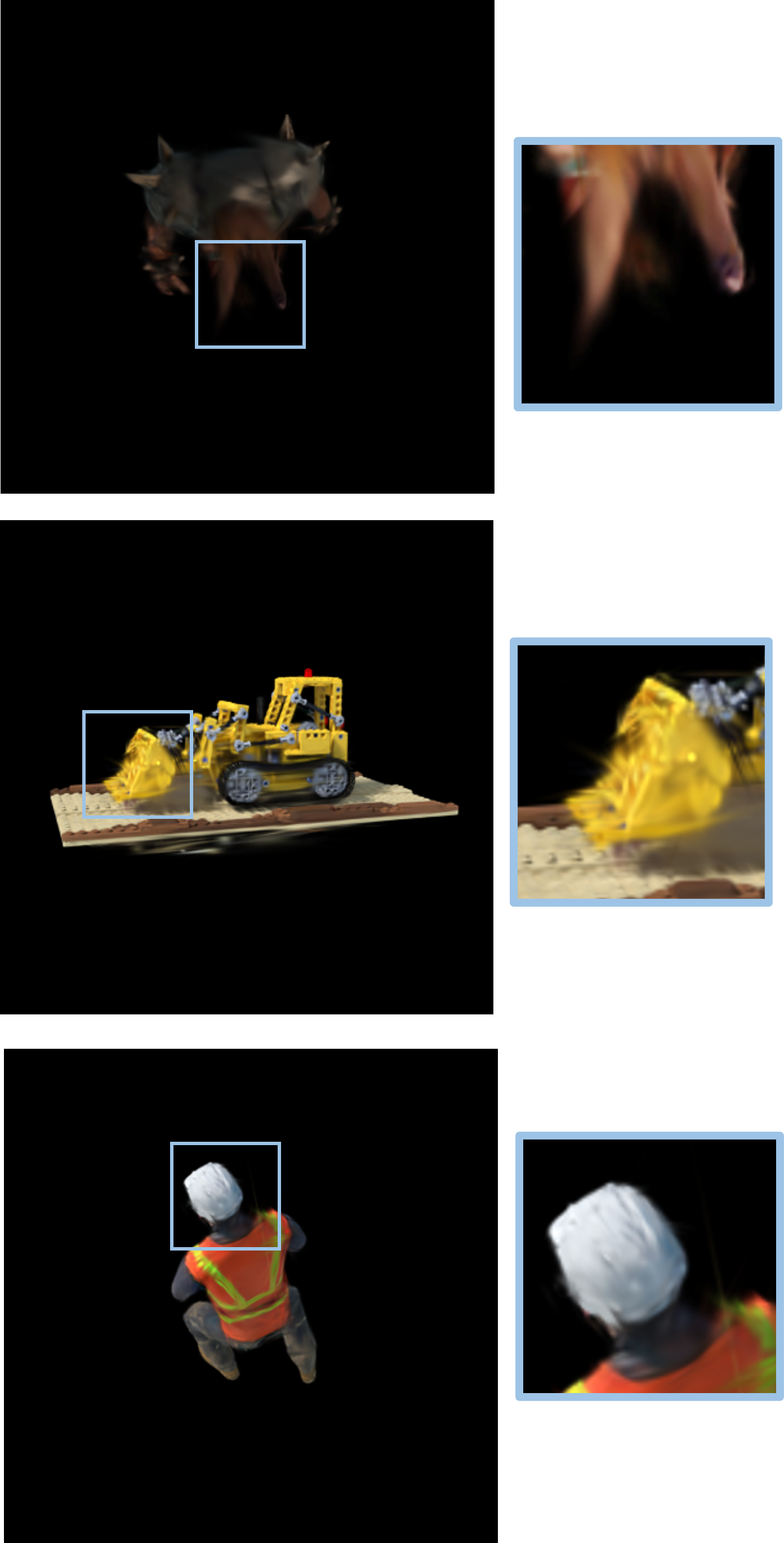}
    \caption{w/o Spatial Edge \\ Loss}
    \label{}
    \end{subfigure}
    \centering
    \begin{subfigure}{0.23\linewidth}
        \includegraphics[width=\linewidth,height=2\linewidth]{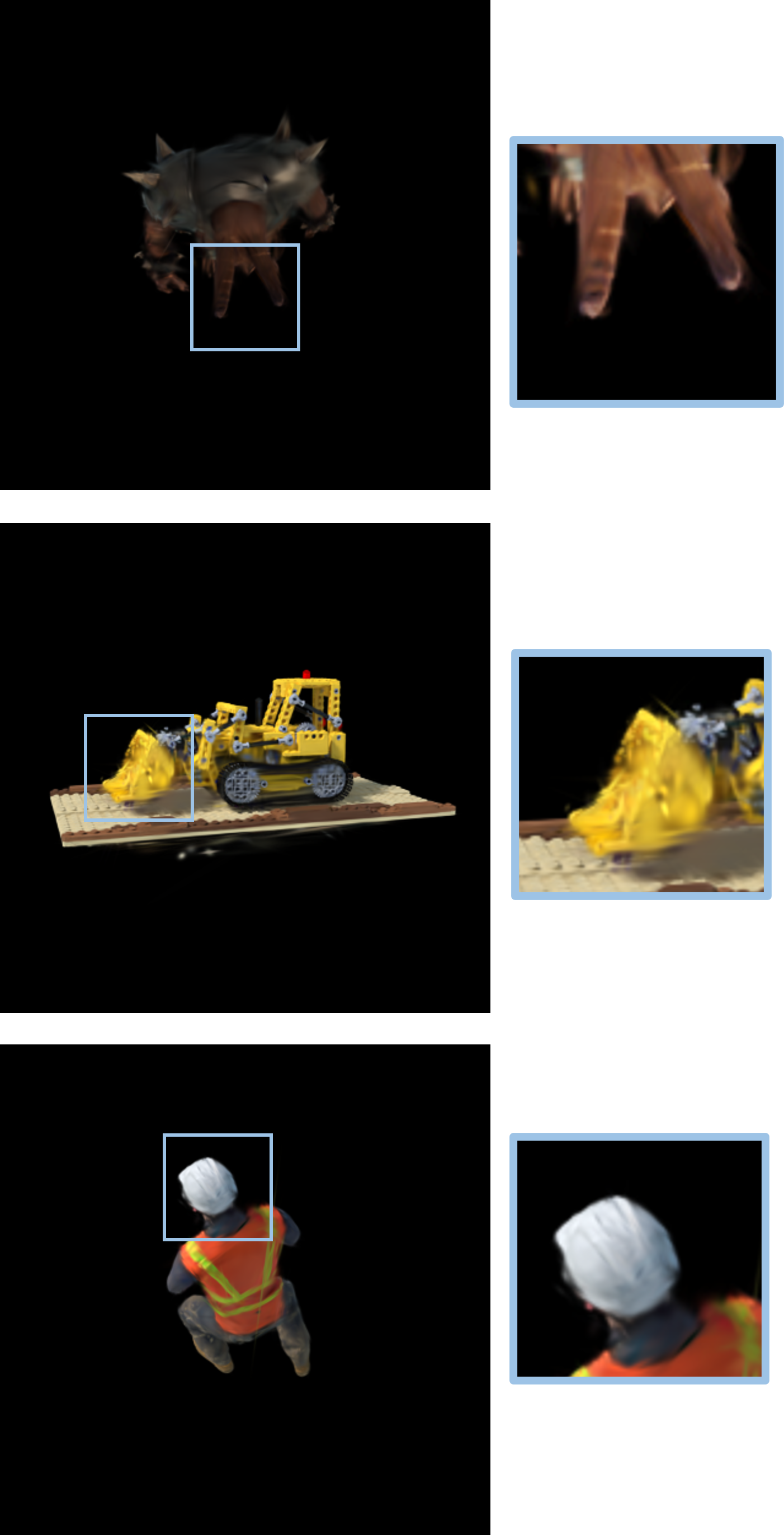}
    \caption{w/o Temporal Splitting \\ strategy}
    \label{}
    \end{subfigure}
    \centering
    \begin{subfigure}{0.23\linewidth}
        \includegraphics[width=\linewidth,height=2\linewidth]{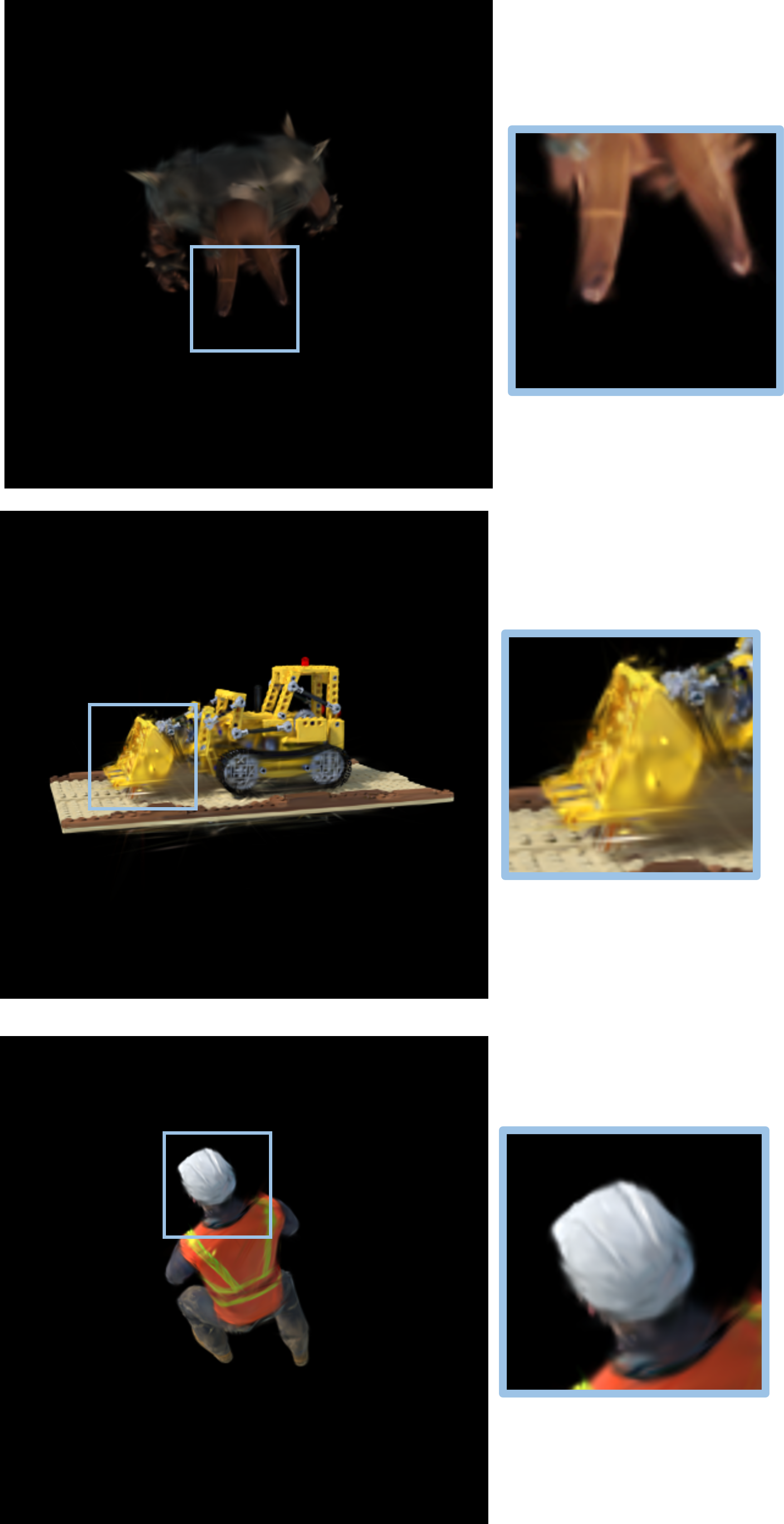}
    \caption{ w/o Temporal Splitting \\ \& Edge Loss}
    \end{subfigure}
    \caption{Visualization of \textit{Lego} scene}
    \label{fig:ablation}
\end{figure*}

To demonstrate the rendering speed advantage of our method, we test the CUDA rendering time of the traditional 4D Gaussian methods and our method under under identical conditions, i.e., the same viewing angle, 100,000 Gaussians. The result in Tab. \ref{tab:speed} shows that our approach significantly improves the rendering speed by 25\% compared to the most recent method~\cite{Duan20244DRotorGS}. And When our method deals with scenes with fixed viewpoints, it can simultaneously render multiple frames in parallel, making our method more efficient in terms of speed. Our method consumes an average of 500 MB VRAM on this dataset for each additional frame during parallel rendering, while the delay only increases about 0.4 ms
\subsection{Ablation and Analysis}

In Tab.~\ref{Tab 3.} and Fig.~\ref{fig:ablation}, we conduct ablation studies on effectiveness of individual designs in our method.
\subsubsection{Flow-Gradient Guided Consistency Loss}
As shown in Tab.~\ref{Tab 3.} (c), our flow-gradient guided consistency loss improves the rendering quality in both PSNR and SSIM and the visual effect of the flow-gradient guided consistency loss is clearly illustrated in Fig.~\ref{fig:ablation}. Specifically, the details in the edge around the scene \textit{Lego}, \textit{Standup}, and \textit{Hellwarrior} are consistently enhanced while the blurs are significantly reduced with our flow-gradient guided consistency loss. This demonstrates that the flow-gradient guided consistency loss helps reduce the artifacts and improve the high-frequency details. Therefore, our method improves the rendering quality especially for the complex scenes like \textit{Lego} in D-NeRF and Plenoptic Video Dataset.
\subsubsection{Temporal Splitting strategy}
As shown in Table \ref{Tab 3.} (b), our temporal split strategy improves the rendering quality of dynamic scenes. By decoupling the classic density control strategy, we successfully reduce spatial and temporal over-reconstruction regions. Furthermore, when combined with flow-gradient guided consistency loss, this strategy significantly alleviates the artifact problem in 4D Gaussian rendering. As illustrated in Figure \ref{fig:ablation}, this results in sharper edges on moving parts such as the bucket of Lego, the antennas of Hellwarrior, and the helmet in the Standup scene.

\section{Conclusion and limitations}
We present Disentangled 4D Gaussian Splatting (Disentangled4DGS), a method that disentangles spatial and temporal components in a 4D Gaussian scene representation and defers temporal processing to avoid expensive 4D matrix operations. By incorporating a flow-gradient guided consistency loss and a temporal splitting strategy, our approach reduces rendering artifacts and achieves real-time performance on an RTX 3090, rendering 1352 × 1014 dynamic scenes at 343 FPS. Compared with previous techniques, Disentangled4DGS delivers superior image quality, stability, and memory efficiency. Future work will investigate further compression of the 4D Gaussian representation and applications in dynamic scene segmentation and generation.

Although our representation method reduces the number of floating-point values and improves storage efficiency, degree of freedom analysis in \textbf{supplementary material} suggests that the 4D Gaussian sphere representation still has further compression potential. Moreover, while our approach offers a clear advantage in rendering speed, it exhibits limitations in rendering quality—particularly when handling extremely sparse inputs such as monocular synthesized reconstruction data. In such scenarios, faithfully capturing dynamic details remains challenging. 

\bibliographystyle{IEEEtran}
\bibliography{IEEE-bib}

\vfill
\begin{IEEEbiography}[{\includegraphics[width=1in,height=1.25in,clip,keepaspectratio]{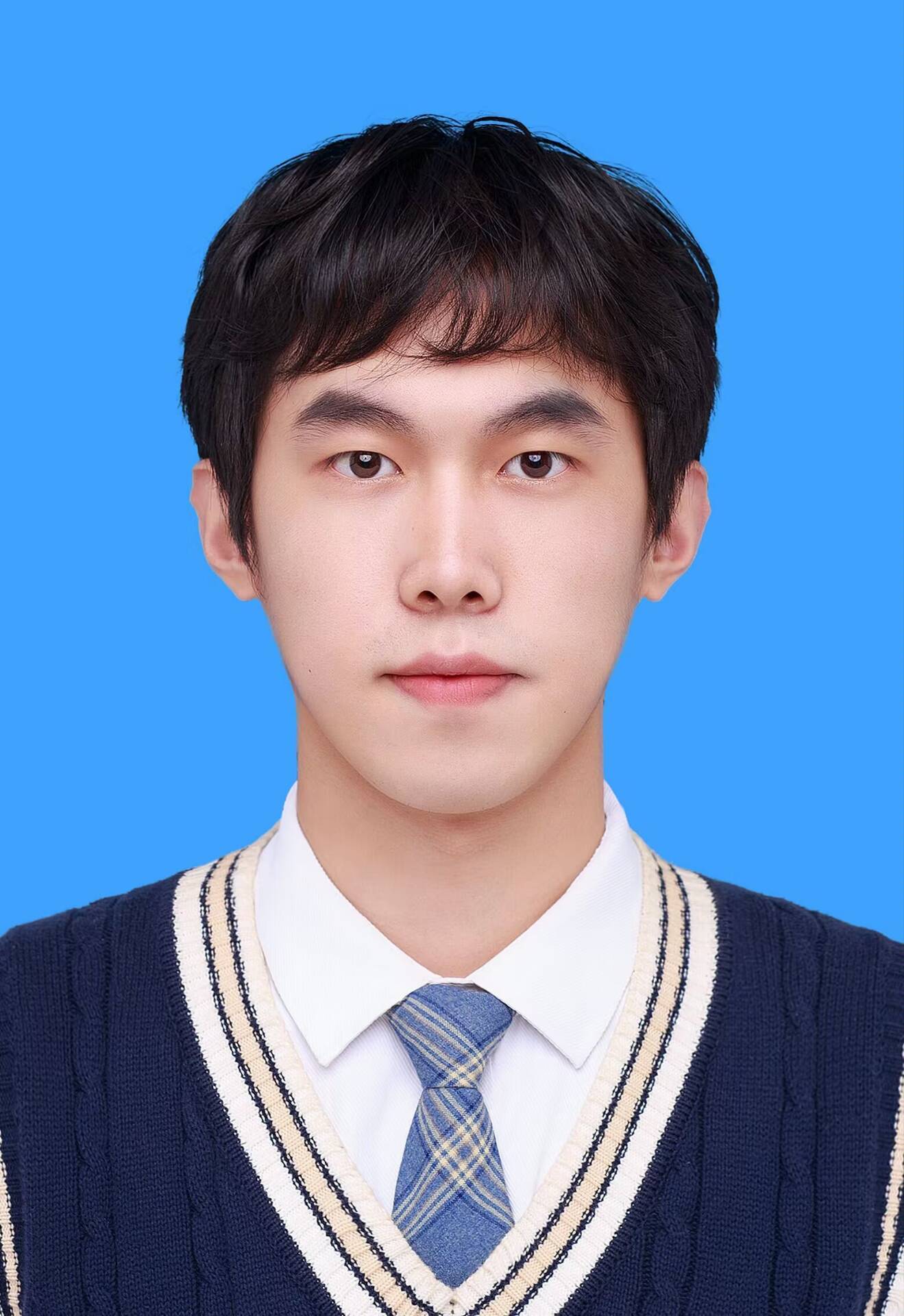}}]{Hao Feng}
Hao Feng received the B.S. degree from the Nanjing
University of Information Science and Technology
(NUIST), Nanjing, China, in 2022. He has entered
the school of computer science, Central China Normal University in 2023
to pursue a master’s degree. His research interests
include video comprehension and 4D reconstruction.
\end{IEEEbiography}
\begin{IEEEbiography}[{\includegraphics[width=1in,height=1.25in,clip,keepaspectratio]{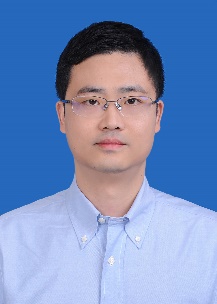}}]{Wei Xie}
is a professor at School of Computer Science, Central China Normal University, Wuhan 430079, Hubei, China. His research interests include image processing, computer vision and deep learning.
\end{IEEEbiography}
\begin{IEEEbiography}[{\includegraphics[width=1in,height=1.25in,clip,keepaspectratio]{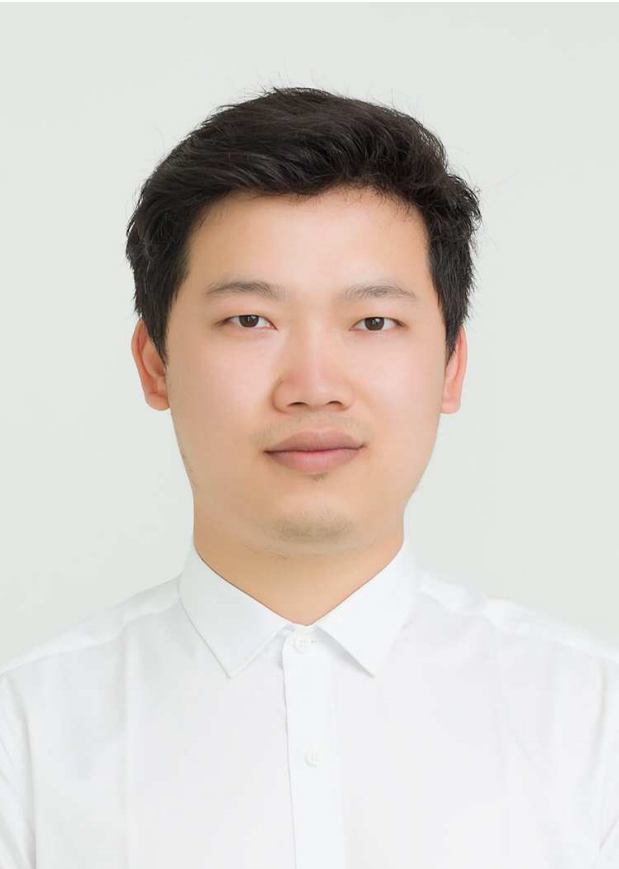}}]{Hao Sun}(Member, IEEE)
received the Ph.D. degree in signal and information processing from the University of Chinese Academy of Sciences, Beijing, China, in 2021.
He is currently a Lecturer with the School of Computer Science, Central China Normal University, Wuhan 430079, Hubei, China.
His current research interests include computer vision, deep learning and hyperspectral image analysis.
\end{IEEEbiography}
\begin{IEEEbiography}[{\includegraphics[width=1in,height=1.25in,clip,keepaspectratio]{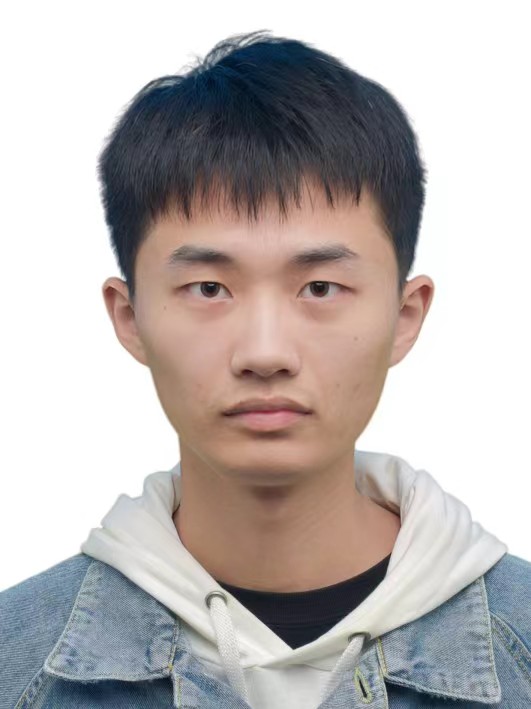}}]{Zhi Zuo} received the B.S. degree from the Nanjing University of Information Science and Technology (NUIST), Nanjing, China, in 2023. He has entered the College of Artificial Intelligence, Nanjing University of Aeronautics and Astronautics in 2023 to pursue a master's degree. His research interests include 3D point cloud analysis and artificial intelligence.
\end{IEEEbiography}
\begin{IEEEbiography}[{\includegraphics[width=1in,height=1.25in,clip,keepaspectratio]{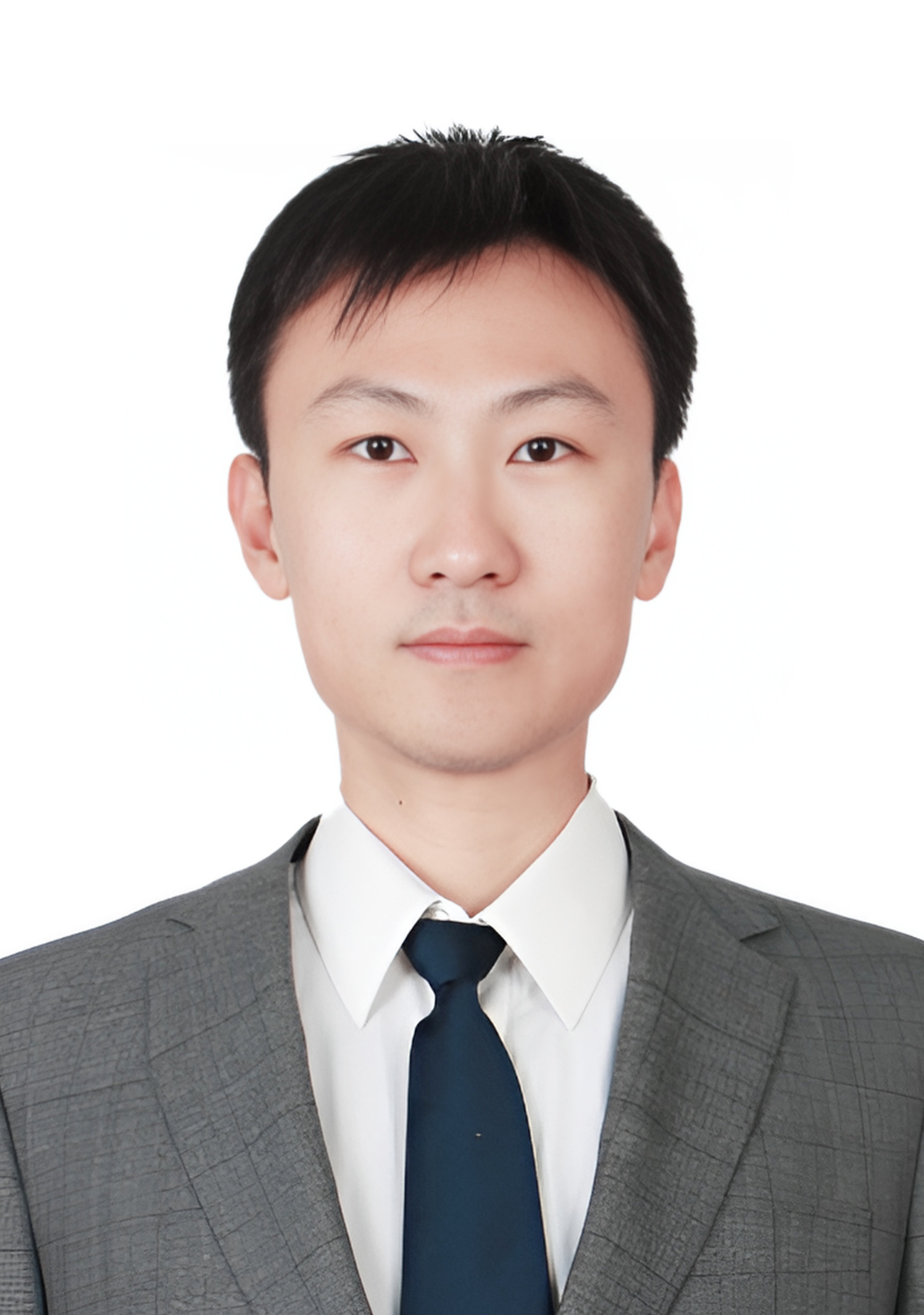}}]{Zhengzhe Liu} is currently an assistant professor at Lingnan University. He received his B.Eng degree in Information Engineering from Shanghai Jiao Tong University, and the M.Phil. and Ph.D. degree in Computer Science and Engineering from The Chinese University of Hong Kong. In 2024, he was a postdoctoral associate at Carnegie Mellon University. His research interests include AIGC, computer graphics, and 3D shape generation
\end{IEEEbiography}

\end{document}